\documentclass{aastex}
\usepackage{apj}
\shorttitle{NICMOS observations of extragalactic Cepheids. I}
\shortauthors{Macri {\it et al.}}
\newlength{\mew}
\def \me {\hskip\mew}
\def \ndw {\hskip 10pt\nodata\hskip 10pt\ }
\def \nd {\nodata}
\def \ng#1{\hbox{NGC$\,$#1}}
\def \ic#1{\hbox{IC$\,$#1}}
\def \ebv {\langle E(B\!-V)\rangle}
\def \evi {\langle E(V\!-I)\rangle}
\def \evh {\langle E(V\!-H)\rangle}
\def \vi {V\!-\!I}
\def \vh {V\!-\!H}
\def \rvi {{\cal R}_{\vi}}
\def \rvh {{\cal R}_{\vh}}
\def \rvl {{\cal R}_{V\!-\!\lambda}}
\begin{document}
\title{NICMOS Observations of Extragalactic Cepheids. I.
Photometry Database and a Test of the Standard Extinction Law$^\dagger$}

\author{L.M.~Macri\altaffilmark{1},
D.Calzetti\altaffilmark{2},
W.L.~Freedman\altaffilmark{3},
B.K.~Gibson\altaffilmark{4},
J.A.~Graham\altaffilmark{5},
J.P.~Huchra\altaffilmark{1},
S.M.G.~Hughes\altaffilmark{6},
B.F.~Madore\altaffilmark{3,7},
J.R.~Mould\altaffilmark{8},
S.E.~Persson\altaffilmark{3} \& 
P.B.~Stetson\altaffilmark{9}}

\altaffiltext{1}{Harvard-Smithsonian Center for Astrophysics, 60 Garden
St. MS 20, Cambridge MA 02138, USA}
\altaffiltext{2}{Space Telescope Science Institute, 3700 San Martin Dr.,
Baltimore, MD 21218, USA}
\altaffiltext{3}{Observatories of the Carnegie Institution of Washington, 
813 Santa Barbara St., Pasadena, CA 91101, USA}
\altaffiltext{4}{Centre for Astrophysics \& Supercomputing, Swinburne 
University of Technology, Hawthorn, Victoria, 3122, Australia}
\altaffiltext{5}{Department of Terrestrial Magnetism, Carnegie Institution of
Washington, 5241 Broad Branch Road NW, Washington, DC 20015, USA}
\altaffiltext{6}{Royal Greenwich Observatory, Madingley Road, Cambridge CB3
0EZ, England}
\altaffiltext{7}{NASA/IPAC Extragalactic Database, California Institute of
Technology, MS 100-22, Pasadena, CA 91125, USA}
\altaffiltext{8}{MSSSO, Research School of Astronomy \&
Astrophysics, Australian National University, Weston Creek, ACT 2611, Australia}
\altaffiltext{9}{Dominion Astrophysical Observatory, Herzberg Institute of
Astrophysics, 5071 West Saanich Road, Victoria, BC V8X 4M6, Canada}

\renewcommand{\thefootnote}{\fnsymbol{footnote}}
\footnotetext[2]{Based on observations with the NASA/ESA {\it Hubble
Space Telescope}, obtained at the Space Telescope Science Institute,
operated by AURA, Inc. under NASA contract No. NAS5-26555.}
\renewcommand{\thefootnote}{\arabic{footnote}}

\begin{abstract}
We present the results of near-infrared observations of extragalactic Cepheids
made with the Near Infrared Camera and Multi-Object Spectrometer on board the
{\it Hubble Space Telescope}. The variables are located in the galaxies
\ic1613, \ic4182, M31, M81, M101, \ng925, \ng1365, \ng2090, \ng3198, \ng3621,
\ng4496A and \ng4536. All fields were observed in the F160W bandpass;
additional images were obtained in the F110W and F205W filters. Photometry was
performed using the DAOPHOT~II/ALLSTAR package.

Self-consistent distance moduli and color excesses were obtained by fitting
Period-Luminosity relations in the $H$, $I$ and $V$ bands.  Our results support
the assumption of a standard reddening law adopted by the HST Key Project on
the Extragalactic Distance Scale. A companion paper will determine true
distance moduli and explore the effects of metallicity on the Cepheid distance
scale.

\end{abstract}


\section{Introduction}
One of the most important legacies of the {\it Hubble Space Telescope} (HST)
will undoubtedly be its revolutionary increase in the number of Cepheid-based
distances to nearby galaxies. Two major projects, the HST Key Project on the
Extragalactic Distance Scale \citep{fr01a} and the HST SN Ia Calibration
Project \citep{sah99}, as well as smaller collaborations have resulted in the
discovery of over 700 Cepheid variables and the determination of distances to
27 galaxies.  This number will continue to grow as the community continues to
take advantage of HST's unparalleled ability to deliver high-resolution images
of the crowded spiral arms where Cepheids are located. The next generation of
instruments to be installed on board HST will increase its sensitivity and
resolution and will extend these observations to larger distances.

Most HST Cepheid observations have concentrated on obtaining data in the $V$
band (typically in eight to thirteen non-aliasing epochs) with some additional
$I$-band data, usually four to eight epochs. Although sparsely sampled, the
$I$-band light curves are important for several reasons. First, they provide
confirmation of the variables as Cepheids, since the $V$ and $I$ light curves
should track each other, with the latter displaying an amplitude of about half
of that of the former. Second, the $V-I$ colors should be consistent with those
of Cepheids in the instability strip. Third, and most important, the $I$-band
data provide the only means of correcting the observed distance modulus $\mu_V$
for extinction through the relation

\vskip -6pt
\begin{equation}
\mu_0 = \mu_V - \rvi \evi
\end{equation}

\noindent{where $\rvi = 2.45$ is the adopted ratio of total-to-selective
extinction for the $V$-band \citep{ccm89} and $\evi$ is the mean color excess
of the Cepheid sample (\S5 contains a calculation of the value of $\rvi$). This
approach makes the true distance modulus quite sensitive to both $\rvi$ and
$\evi$. A better procedure \citep{fm90} involves fitting a standard extinction
curve through observed distance moduli at several wavelengths (usually $B$,
$V$, $R$ and $I$) and extrapolating the fit to $\lambda^{-1} = 0$. This
approach is less susceptible to uncertainties in the individual observed
distance moduli, and can also be used to test the assumed reddening law.}

As the HST Key Project on the Extragalactic Distance Scale started its final
cycle of observations, a subset of the team joined forces with colleagues
outside the group to extend the work further into the infrared. Our aim was to
perform random-phase near-IR photometry of a sample of Cepheids in fourteen
galaxies and to combine the new and existing observed distance moduli. Such a
data set would allow us to improve the derivation of true distance moduli,
check the assumptions behind Equation (1), and perhaps explore the effects of
metallicity on the Cepheid distance scale.

The paper is organized as follows: \S 2 describes the observations and the data
reduction pipeline; \S 3 delineates the steps followed to obtain accurate and
precise photometry in our fields; \S 4 presents the Cepheid sample,
period-luminosity relations, and observed distance moduli; \S 5 discusses our
results.

\section{Observations and Data Reduction}

\subsection{Observations}

The HST Near Infrared Camera and Multi-Object Spectrograph (NICMOS) instrument
\citep{tho93}, with its high spatial resolution and low thermal background, was
uniquely suited to carry out the observations required by our program. NICMOS
contains three cameras (named 1, 2 and 3) which illuminate Rockwell $256\times
256$ HgCdTe arrays. The cameras have different pixel scales (0\farcs 043,
0\farcs 076 and 0\farcs 2, respectively), resulting in fields-of-view of
11\arcsec, 19\arcsec\ and 51\arcsec, respectively. Since Cepheids are scattered
throughout the spiral arms of the target galaxies, we chose Camera 2 (hereafter
NIC2) as it provided the best trade-off between resolution and coverage.

The fourteen galaxies selected for this study are listed in Table 1. The
selection of specific fields within each galaxy was based on known positions
and periods of Cepheids, in order to maximize the number of variables and our
coverage of the Period-Luminosity plane. We observed two fields in M101,
matching those observed by \citet{ste98} and \citet{kel96}. These will be
hereafter referred to as ``M101-Inner'' and ``M101-Outer.''

The observations followed the standard {SPIRAL-DITH} pattern, with two to
four pointings depending on the field and filter used. Exposure times for each
pointing ranged from 16s for M31 to 640s for the most distant galaxies. Some of
the latter fields were imaged multiple times in order to increase measurement
precision.

\subsection{Data reduction}

Once NICMOS had been installed on board HST, several instrument characteristics
were discovered. One is a variable additive bias (called ``pedestal'')
introduced during array reset, which has a different amplitude in each of the
four quadrants that make up the array.  Because HgCdTe arrays do not have
overscan regions, it is impossible to automatically remove this effect in the
STScI pipeline. Therefore, the bias offset is modulated by the flat field and
appears as an inverse flat-field pattern in the final image.  A second
characteristic of the detectors is a noiseless signal gradient (called
``shading''), which is a temperature- and pixel- dependent bias that changes in
the direction of pixel clocking during read-out. This presented a problem
because the original implementation of the STScI processing pipeline did not
use temperature-dependent darks.  The combination of ``pedestal'' and
``shading,'' resulted in images with prominent spurious features. The level of
photometric precision required by our program made it necessary to remove these
instrumental effects. We did so by retrieving the raw frames from the STScI
Archive and reprocessing them as described below.

Our reprocessing of the data was performed using the NICMOS pipeline present in
IRAF\footnote[1]{IRAF is distributed by the National Optical Astronomy
Observatories, which are operated by AURA, Inc., under cooperative agreement
with the National Science Foundation.}/STSDAS with some modifications.  First,
temperature-dependent dark frames were generated and used when running the
first part of the pipeline (a program called {calnicA}). This removed the
``shading'' effect.

The ``pedestal'' effect was corrected using a program outside of the standard
pipeline, courtesy of R. van der Marel (STScI). The program read in an image
created by {calnicA}, removed the flat-fielding imposed by it, and executed
a loop to identify the pedestal. The pedestal is in fact measured by exploiting
the property that a flat-fielded bias imparts fluctuations on the background of
the final image. The fluctuations reach a minimum when the pedestal is removed
completely.  Once a robust minimum was found in each of the quadrants, the
best-fit pedestals were removed.  Lastly, the image was flat-fielded and
written to disk.

Having obtained images with proper zero, dark and flat-field corrections, the
second part of the pipeline ({ calnicB}) was run to combine the dithered
pointings of each target field and produce a final mosaic. In addition, we used
the {dither} package to produce higher-resolution mosaics of the M101
fields in the F160W band. This was possible thanks to the existence of four
pointings per field in that galaxy.

\section{Photometry}

\subsection{Technique}

The mosaics were analyzed with the {DAOPHOT~II/ ALLSTAR} software package
\citep{ste94}.  Objects were detected with the {FIND} routine set to a
threshold of $3\sigma$ above sky, and aperture photometry was carried out with
the {PHOT} routine, using different apertures and sky annuli for each
filter (see below for details).  Point-spread functions were determined by the
{ PSF} routine for each field using bright, isolated stars present in the
frame. After an initial {ALLSTAR} run, the star-subtracted frame was put
through the {FIND} algorithm once more to pick up any additional $3\sigma$
objects. Object lists were merged and {ALLSTAR} was run one last time on
the original image.

PSF magnitudes were brought onto a consistent aperture magnitude system using
{ DAOGROW} (see \citealt{ste90} for details; only a short summary is
presented here). Aperture photometry was obtained for PSF stars in each field
($\sim 450$ in total) using a monotonically increasing set of radii. {
DAOGROW} then solved for a function representing the ``growth curve'', i.e.,
the change in aperture magnitude as a function of radius. The cumulative
growth curves for the F110W, F160W and F205W bands are shown in Figure
1. Other programs in the {DAOPHOT} suite used this information to
transform PSF magnitudes to aperture magnitudes for objects of interest in
each frame.

\subsection{Absolute photometric calibrations}

The NICMOS primary standards are G191-B2B, a white dwarf, and P330-E, a solar
analog. These two stars provide absolute calibration in the white dwarf and
solar analog scales, respectively.  Since Cepheids have colors similar to those
of solar analog stars, we used P330-E for the determination of magnitude
zeropoints. We used NICMOS observations of G191-B2B, P330-E, and P117-D
(another solar analog standard) for the determination of color terms.
Ground-based $JHK$ photometry for NICMOS standards comes from \citet{per98}.

\subsubsection{Magnitude zeropoints}

Our definition of $J$, $H$ and $K$ zeropoints is based on different apertures
and sky annuli for each band, to match the noticeable increase in FWHM as a
function of wavelength. Table 2 lists our choices of aperture radius and inner
and outer sky annuli for each of the three bandpasses. In the case of the
drizzled M101 F160W images, which have twice the spatial resolution, all radii
were increased by a factor of two in pixel units so they would subtend the
same angular size.

Marcia Rieke kindly provided us with synthetic (TinyTim) stellar images as well
as NICMOS observations of P330-E, which we used to derive the magnitude
zeropoints for our choices of aperture and sky annuli. First, we computed
the ratio of TinyTim counts to observed NICMOS count rates for P330-E as a
function of aperture radius. We found this ratio to be constant, at the 0.05\%,
0.03\%, and 1.77\% levels for F110W, F160W and F205W, respectively, over a
large range in radius (5--25 pixels). This confirmed that the TinyTim image
was a good representation of the actual system PSF.

We then re-scaled the TinyTim image (in arbitrary units) to match the actual
mean observed count rate of P330-E; this produced a synthetic image of P330-E
that could be used to perform aperture measurements on a ``perfect'' image
free of defects, cosmic rays, or any other source of scatter found in real
images.  Next, we ran DAOPHOT's {PHOT} routine on the synthetic P330-E
images using the same aperture and sky annuli as for the Cepheid photometry.
DAOPHOT quoted measurement errors of 0.004 mag for these magnitudes.  Lastly,
we combined the \citet{per98} standard magnitudes and the DAOPHOT NICMOS
instrumental magnitudes for P330-E to arrive at our magnitude zeropoints,
listed in Table 2.

\subsubsection{Color terms}

Since the NICMOS filters are not exact matches to the standard filters, color
term corrections had to be determined. Synthetic spectra of the NICMOS
standards were created based on Kurucz' latest solar-abundance models. These
model spectra were convolved with two sets of transmission curves: one
contained the NICMOS filter responses plus the quantum efficiency of camera 2,
while the other was based on standard filter responses plus an atmospheric
transmission curve.  This allowed us to predict F110W, F160W, F205W, $J$, $H$
and $K$ magnitudes for the NICMOS standards. We then compared our results with
published values \citep{per98} and found negligible offsets of
$0.002\pm0.003$~mag for $J$ and $H$ and a small offset of $0.022\pm0.001$~mag
for $K$.

We used the same model atmospheres and transmission curves described above to
generate synthetic spectra for a variety of spectral types (F, G and K) and
luminosity classes (I and V). We compared the values of $\langle F110W-J
\rangle$ and $\langle F160W-H \rangle$ as a function of $\langle
F110W-F160W\rangle$ as well as the values of $\langle F205W-K\rangle$ as a
function of $\langle F160W-F205W\rangle$.  We found

\hfill

\begin{eqnarray}
\langle F110W-J \rangle &=& 0.013\pm0.006 +  \nonumber \\
              0.315 &\pm&0.011\ \langle F110W-F160W\rangle \\ 
\langle F160W-H \rangle &=& 0.003\pm0.002 +  \nonumber \\
              0.071 &\pm&0.005\ \langle F110W-F160W\rangle \\
\langle F205W-K \rangle &=& 0.053\pm0.008 -  \nonumber \\
              0.556 &\pm&0.085\ \langle F160W-F205W\rangle
\end{eqnarray}

\noindent{where $\langle F110W-F160W\rangle$ and $\langle F160W-F205W\rangle$
are the mean instrumental colors of the star to be corrected. These formulae
are suitable for correcting our data since our F110W, F160W and F205W
observations were taken within minutes of each other.}

The mean values of the corrections were $0.23\pm0.11$~mag for $\langle
F110W-J\rangle$, $0.05\pm0.02$~mag for $\langle F160W-H\rangle$ and
$-0.04\pm0.08$~mag for $\langle F205W-K\rangle$. Note that an exact correction
for $\langle F160W-H\rangle$ was only applied to the stars in the \ic1613,
M31, M101-Inner and M101-Outer fields. The other fields were observed only in
F160W and therefore only an average H-band correction of $0.03$~mag could be
applied, based on a mean Cepheid $\langle F110W-F160W\rangle$ color of
$0.46$~mag.

\subsection{Photometric recovery tests}

The dense nature of most of our fields makes it difficult to obtain accurate
values of the local sky around each object and to perform unbiased magnitude
measurements. This effect is commonly referred to as ``crowding''. In order to
characterize its impact on our measurements, we injected artificial stars into
each field and compared their input magnitudes with the recovered values. We
used the point-spread functions derived for each field to generate the
artificial stars, which were placed randomly across each field.  The objects
spanned a magnitude range including that encompassed by the variables. In the
case of the M101 fields, we injected artificial stars into the
original-resolution mosaics as well as the higher-resolution ones created by
drizzling. We re-ran our photometry programs on the new images and searched
the new star lists to locate the artificial stars.

The results of the tests are summarized in Table 3 and displayed in Figures
2a-b. Figure 2a contains plots of the difference between input and recovered
magnitudes as a function of magnitude for each field. Figure 2b shows the
strong correlation that exists between the crowding bias and the stellar
density of each field. The effect ranges from 0.01~mag for the least crowded
fields to 0.09~mag for the denser ones. All magnitudes were corrected for
this effect.

In addition to these ``crowding'' tests, we also undertook simulations to
estimate the contamination of Cepheid magnitudes by unresolved nearby
stars. These ``blending'' tests are presented in \S 5.

\subsection{Photometry checks}

We performed several internal and external photometry checks to ensure the
accuracy and precision of our magnitudes. We tested our aperture correction
technique by comparing our corrected magnitudes against ``standard'' aperture
magnitudes for stars in the IC 1613 fields. We found no significant difference
($<0.01$~mag) between the two sets. We also tested the repeatability of our PSF
photometry by comparing magnitudes of objects that appeared twice in our data
set, due to some overlap between different fields in M81 and M101. We found
that the magnitude differences were consistent with the reported uncertainties.

We also performed an external photometry check by comparing HST and
ground-based photometry of bright, isolated stars in our \ic1613 fields. The
ground-based photometry was obtained at the Las Campanas 2.5-m du Pont
telescope using its infrared camera \citep{per92} over fourteen nights between
November 1993 and November 1996. Photometry was conducted using {
DAOPHOT~II/ALLSTAR} and {DAOGROW} (as described in \S3.1) on 19 stars
common to our ground-based and HST images. Table 4 lists their magnitudes and
Figure 3 shows a comparison of the two systems; the mean offset is
$0.011\pm0.061$~mag, in the sense that the HST magnitudes are marginally
brighter. Unfortunately, there was no published $J$- or $K$-band photometry
available for stars in any of our fields, so we were unable to check our
transformation of NICMOS F110W and F205W magnitudes.

\section{The Cepheid sample}

\subsection{Sample selection and identification}

As described in \S2.1, we targeted specific fields within each galaxy in order
to maximize the number of variables and our coverage of the Period-Luminosity
plane. We selected the variables in each galaxy based on published catalogs,
applying the following selection criteria: i) existence of both $V$ and $I$
photometry; ii) range in color of $0.5 < \vi < 1.75$; iii) periods between 10
days and the width of the observing window (applicable to Cepheids discovered
with HST). Our fields contained 93 variables that met these criteria.

Cepheids in M31 and IC1613 were identified by visual inspection, using finding
charts created from ground-based images. These fields are sparse enough that
identifications did not present a problem, and twelve variables were
located. In the case of the other galaxies, identifications followed a more
rigorous process. First, the FITS header coordinates for the center of the
mosaic were used to obtain a rough alignment and rotation relative to an
optical image (from WFPC2 in most cases).  Next, bright stars present in both
the optical and the near-IR images were identified and used as input to the
IRAF task {geomap} to determine the geometric transformations between the
images. Lastly, the task {geoxytran} was used to predict the coordinates
of 81 variables.

The DAOPHOT star lists generated in \S3.1 were used to locate the object
nearest to the predicted position of the variables. In general, counterparts
were found within one pixel of their predicted location. Figure 4 shows the
distribution of differences between the predicted and actual positions. Based
on this figure, we decided to reject any candidate located more than 1.5
pixels (0\farcs11) away from its predicted position. This process resulted in
the rejection of 11 possible counterparts.

In order to further discriminate between real counterparts and field
contaminants, we plotted $\vh$ vs. $\vi$ colors for all remaining candidates
(Figure 5a). The Cepheids follow a vector that is a combination of two closely
degenerate quantities: the reddening trajectory and the color-color relation
for these bands. Several objects deviate significantly from the rest of the
sample; we suspect these are variables which are blended with unresolved red
or blue companions. We performed a least-squares fit to the sample, using a
fixed slope of $\vh/\vi = 1.71$ (the average of the reddening and color-color
slopes). The fit is shown in Fig. 5a as a solid line, while the dashed lines
correspond to twice the {\it r.m.s.} deviation, or 0.46~mag. We rejected twelve
possible counterparts that fell outside of the dashed boundaries. 

Figure 5b shows a histogram of the deviations from the best-fit line. The
asymmetric distribution of the outliers is to be expected, since we are more
likely to detect a Cepheid that is blended with a red (i.e., IR-bright) field
star than with a blue (i.e., IR-faint) one. Note that this color-color
rejection process is insensitive to blends of Cepheids with stars of similar
colors, a point to which we will return later.

In conclusion, our final sample consists of 70 variables (93 original
candidates -- 11 astrometric rejections -- 12 color-color rejections). Finding
charts for fields containing at least one variable are shown in Figures 6a-f,
while Figures 7a-b contain close-up views of each object.  Table 5 presents
periods and magnitudes for the final Cepheid sample. We include in this Table
the previously-published optical magnitudes of the variables. There are minor
variations in the $V$ and $I$ zeropoints used in the different sources of
optical photometry, reflecting the evolution in our knowledge of the HST
calibration from 1994 to the present (see \citealt{mou00} for details).  For
our target galaxies, the mean difference between the various calibrations
used in the published papers and the current calibration \citep{st98} amounts
to $-0.03\pm0.03$~mag in V, $-0.05\pm0.04$~mag in I and $+0.02\pm0.02$~mag in
$\vi$.

\subsection{Period-Luminosity relations}

The method used to derive observed distance moduli is the same as that used
by the HST Key Project on the Extragalactic Distance Scale \citep[see][for
details]{fr01a}. It is based on the Period-Luminosity relations of
individually de-reddened LMC Cepheids from \citet{ud99} ($V$ and $I$) and
\citet{per01} ($J$, $H$ and $K$), scaled to an assumed true distance
modulus of $\mu_{0,{\rm LMC}}$ = 18.50$\pm$0.10~mag (total uncertainty). The
relations are:

\vskip -6pt
\begin{eqnarray}
M_V&=&-2.76(\pm0.03)\left[\log{\rm P}-1\right] -4.22(\pm0.02), \\	     
M_I&=&-2.96(\pm0.02)\left[\log{\rm P}-1\right] -4.90(\pm0.01), \\
M_H&=&-3.23(\pm0.04)\left[\log{\rm P}-1\right] -5.66(\pm0.05), \\
M_J&=&-3.15(\pm0.05)\left[\log{\rm P}-1\right] -5.32(\pm0.06), \\
M_K&=&-3.26(\pm0.04)\left[\log{\rm P}-1\right] -5.73(\pm0.05).   
\end{eqnarray}
								    
In fitting the data from each field and filter, we fix the slope to the one
given in the corresponding equation and obtain a magnitude shift by
minimizing the unweighted {\it rms} dispersion. The resulting magnitude
shifts are converted to observed distance moduli by subtracting the relevant
magnitude zeropoint. 

\subsection{Observed distance moduli}

Period-Luminosity relations were constructed for each field and filter using
the data listed in Table 5 and fitted using Equations 5-9. Figures 8a-f show
our near-IR P-L relations, while Figures 9a-e present the optical ones. In each
panel, the solid line represents the results of the fitting process described
in \S4.2 while the dashed lines indicate the {\it rms} uncertainty of the
fit. Fit results are displayed in each panel and listed in Tables 6 and 7.

We also tabulate in Table 6 published distance moduli for these galaxies
\citep[mostly from Table 4 of][]{fr01a}, determined from substantially larger
samples of variables and using the \citet{st98} zeropoints. The optical
distance moduli determined from our smaller samples should {\it not} take
precedence over the above values. The optimum combination of the optical and
infrared results will appear in \citet{fr01b}.

\section{Blending effects in M101 Inner}

As mentioned in the introduction, one of the motivations of this project was
to further study the metallicity dependence of the Cepheid Period-Luminosity
relation. Our data can contribute to these studies on two ways: first, a
global test of the metallicity dependence can be performed by analyzing the
apparent distance moduli of all galaxies; second, a differential test of the
effect can be performed by analyzing the distance moduli to two regions of
the same galaxy, provided they differ significantly in abundance. The first
approach was undertaken by \citet{ko97}, while the second one was followed by
\citet{fm90} in M31 and by \citet{ke98} in M101. The differential test is a
challenging one, because the Inner field of M101 deviates substantially from
other Key Project fields in terms of surface brightness and stellar density.

Our near-infrared distance moduli to the inner and outer fields in M101
exhibit large differences: $\Delta\mu_H = 0.46\pm0.12$~mag and $\Delta\mu_J =
0.37\pm0.12$~mag. If taken at face value, they imply a very large metallicity
dependence, of the order of 0.6~mag/dex. However, other observational effects
could be contributing to the observed differences in distance moduli.  One of
them is ``blending'', or the contamination of Cepheid fluxes by nearby stars,
not physically associated with the variables, that fall within the NICMOS
seeing disk and cannot be resolved. This effect has been the subject of a
recent investigation in the optical by \citet{moc00}.

One way to characterize the effect of blending on our distance determination
to M101-Inner is to move nearby, well-resolved fields to the distance of M101,
re-observe Cepheids in these fields, and compare the resulting magnitudes with
the ones obtained from the original images. Our program contains observations
of two suitable galaxies: M31 and M81. The fields observed in these galaxies
show similar stellar densities and mean nearest-neighbor distances between
Cepheids when compared to our M101-Inner fields. We used our two M81 fields
as well as fourteen of our M31 fields, which were located in Fields I and III
of \citet{bs65}.

We started by collecting the positions and magnitudes reported by ALLSTAR for
all objects in our input fields. The separation between stars were reduced by
the ratio of distances between the input galaxies and the M101-Outer field
($D(M101/M31)=10.1$ and $D(M101/M81)=2.2$). The input magnitudes were
corrected for the exposure time of the original frames and distance to the
input galaxies, and then modified to reflect the distance of M101 and the
exposure time of the M101 frames. The stars were added to blank frames using
the ADDSTAR routine found in DAOPHOT, which takes into account the properties
of the detector and the PSF, as well as photon statistics. The artificial
fields are shown in Figure 10 and compare favorably with the actual
M101-Inner images shown in Figures 6a-f.

Once the artificial fields had been generated, they were photometered in
exactly the same way as our real data. To identify the variables in our
artificial frames, we used the input positions as the equivalent of the
astrometric information available for the real data, and searched for the
objects nearest to those positions, subject to the same 1.5-pixel rejection
criterion from \S4.1. All Cepheids that were recovered were located at
distances smaller than the rejection limit.

Figure 11 shows the P-L relations obtained from the simulation, compared with
the original input data. Seven long period Cepheids ($P > 20$~d) exhibit
changes in magnitude of order 0.1-0.2~mag, most of them being in the expected
direction (i.e., towards brighter magnitudes). In the case of the eight
Cepheids with short periods ($P < 20$~d), one was not recovered, four exhibited
large variations (which would have resulted in their rejection based on
color-color criteria), and three had very modest changes in magnitude.

The resulting distance moduli are smaller than the input ones by 0-0.2 mag,
depending on the period cutoff applied to these small samples. The {\it rms}
scatter of the relations do not increase significantly, once the short-period
outliers are removed from the fits. We therefore conclude that a substantial
fraction of the difference in distance moduli between M101-Inner and
M101-Outer could be due to blending. This prevents us from performing a
differential determination of the metallicity effect; however, a global test
can still be performed using the other galaxies we have observed.

\section{Consistency of reddening determinations}

Another goal of this project is to test whether the mean $\vi$ color excess,
$\evi$, is an appropriate indicator of total extinction and whether it can be
used to obtain true distance moduli. One could claim that the range in
wavelength between these two bands is too small to allow a good extrapolation
to $\lambda^{-1}=0$. There is also no guarantee that a ``standard'' value of
$\rvi$ is applicable to other galaxies. If $\evi$ is indeed a good indicator
of reddening, and if a standard reddening law \citep{ccm89} applies to other
galaxies, then one would expect $\evi$ and $\evh$ to be strongly correlated
and to follow the slope predicted by the standard reddening law.

\subsection{Predicted relation between $\evi$ and $\evh$}

The mean $\vi$ and $\vh$ color excesses of a Cepheid sample
are related by

\begin{equation}
\evh = \frac{\rvi}{\rvh}\ \evi .
\end{equation}

Furthermore, the value of $\rvl$ ($\lambda$ denotes the bandpass
of interest) can be calculated using the following relation:

\begin{equation}
\frac{1}{\rvl} = 1 - \frac{A_\lambda}{A_V},
\end{equation}

\vskip 6pt

\noindent{where the ratio $A_\lambda/A_V$ is defined in Equation (1) of \citet{ccm89}
as}

\begin{equation}
A_\lambda / A_V = a(x) + b(x)/R_V.
\end{equation}

In turn, $a(x)$ and $b(x)$ can be calculated using Equation (2) of
\citet{ccm89}; $x$ is the inverse of the central wavelength of the band of
choice. The value of $R_V \equiv A_V/\ebv$ suitable for Cepheids and stars of
similar colors is 3.3. Lastly, Figure 3 of \citet{ccm89} can be used to
estimate the size of the uncertainty in $A_\lambda/A_V$.

In our case, we want to calculate the values of $\rvi$ and $\rvh$. The Cousins
$I$ filter has $x=1.23\micron^{-1}$, so $a(x)=0.77$ and $b(x)=-0.59$
\citep{kel96}. Thus, $A_I = 0.59\pm0.03$ and $\rvi = 2.45\pm0.10$. The $H$
filter has $x=0.63\micron^{-1}$, so $a(x)=0.27$ and $b(x)=-0.25$
\citep{ccm89}. Thus, $A_H = 0.19\pm0.03$ and $\rvh = 1.24\pm0.20$. Therefore,
the predicted ratio of $\evh$ to $\evi$ is

\begin{eqnarray}
\evh & = & \frac{2.45\pm0.10}{1.24\pm0.20}\ \evi \nonumber \\
     & = & 1.98\pm0.16\ \evi.
\end{eqnarray}

\subsection{Observed relation}

Mean $\vi$ and $\vh$ color excesses were calculated following the methodology of
the HST Key Project on the Extragalactic Distance Scale \citep{fr01a}. We used
Period-Color relations based on the Period-Luminosity relations from 
Equations 5-7:

\vskip -6pt
\begin{eqnarray}
V-I&=&-0.20(\pm0.04)\left[\log{\rm P}-1\right] +0.68(\pm0.02), \\	     
V-H&=&-0.47(\pm0.05)\left[\log{\rm P}-1\right] +1.44(\pm0.05).
\end{eqnarray}

\vskip 6pt

The mean color excess of a field was calculated by averaging over the
individual color excesses of the variables in that field. The total scatter
about the average value of the color excess in each field was used to
determine the quoted uncertainty on the mean. The values of $\evi$ for
\ic4182, M101 (Inner \& Outer), \ng925, \ng1365, \ng2090, \ng3621, \ng4496A
and \ng4536 were corrected by +0.02~mag to bring them into the \citet{st98}
photometric system (the other galaxies have ground-based or WF/PC V and I
values and need not be corrected).

The mean values of $\evi$ and $\evh$ are listed in Table 8 and plotted in
Figure 12. A least-squares fit to the data yields

\vskip -6pt
\begin{equation}
\evh = 2.02\pm0.22 \evi - 0.05\pm0.05.
\end{equation}

The agreement between the predicted and observed ratio of $\evh$ to 
$\evi$, and the fact that the fit to the data goes through $(0,0)$ within the
errors, supports the assumption of a standard reddening law in the fields we
have studied.

One data point, corresponding to \ng3198 and plotted with an open circle, was
excluded from the fit because it lies $4\sigma$ away from the relation defined
by all other points. This field contains only three Cepheids, two of which
barely passed the color-color rejection test and are probably contaminated by
red companions, therefore yielding an abnormally high value of $\evh$. 

It is interesting to note that the Cepheids present in the M101-Inner field
exhibit the same correlation between $\evi$ and $\evh$ as the other fields.
This could imply that, on average, the contamination due to ``blending'' in
that field has not introduced a significant change in the color of the
Cepheids. \citet{moc01} have found a similar effect among Cepheids in the
inner regions of M33.

\section{Summary}

We have obtained near-infrared photometry for a sample of 70 extragalactic
Cepheid variables located in thirteen galaxies ranging in distance from the
Local Group to the Virgo and Fornax Clusters. We have combined our magnitudes
with existing optical data to derive self-consistent Period-Luminosity
relations.

Cepheids in the inner field of M101 appear to be severely contaminated by
unresolved blends with nearby stars, thereby affecting our ability to perform
a differential test of the dependence on metallicity of the Cepheid
Period-Luminosity relation.

An analysis of mean color excesses of our sample supports the assumption of a
standard reddening law by the HST Key Project on the Extragalactic Distance
Scale in their derivation of true distance moduli.

\section{Acknowledgments}

We would like to thank Eddie Bergeron of STScI for kindly providing us with
temperature-dependent dark frames. We would also like to thank Roland van der
Marel of STScI for kindly providing us with the software used to remove the
pedestal effect. This project was supported by NASA through grant No. GO-07849.

\clearpage

\begin{deluxetable}{lcrr}
\tablenum{1}
\tablecolumns{5}
\tablewidth{0pt}
\tablecaption{Log of observations}
\tablehead{\colhead{Galaxy} & \colhead{Number}    & \colhead{Filter} &
\colhead{Exp. time} \\ \colhead{name}   & \colhead{of fields} & &
\colhead{/field (s)}}
\startdata
IC1613 &  6 & F110W &    32 \\
       &    & F160W &   128 \\\tableline
IC4182 &  4 & F160W &  1152 \\\tableline
M31    & 18 & F110W &    46 \\
       &    & F160W &    89 \\
       &    & F205W &   285 \\\tableline
M81    &  2 & F160W &  1280 \\\tableline
M101   &  8 & F110W &   512 \\
       &    & F160W &  2048 \\\tableline
N0925  &  2 & F160W &  2560 \\\tableline
N1365  &  2 & F160W &  5120 \\\tableline
N2090  &  2 & F160W &  2560 \\\tableline
N2403  &  2 & F160W &  1280 \\\tableline
N3198  &  2 & F160W &  2560 \\\tableline
N3621  &  1 & F160W &  5120 \\\tableline
N4496A &  1 & F160W & 10240 \\\tableline
N4536  &  1 & F160W & 10240 \\\tableline
N5253  &  5 & F160W &   973 \\

\enddata
\end{deluxetable}

\begin{deluxetable}{crllll}
\tablenum{2}
\tablecolumns{6}
\tablewidth{0pt}
\tablecaption{Photometric systems used in the project}
\tablehead{\colhead{Filter} & \multicolumn{2}{c}{Aperture} & \multicolumn{2}{c}
{Sky annulus} & \colhead{Magnitude} \\
\colhead{} & \colhead{pix} & \colhead{\arcsec} & \colhead{pix} &
\colhead{\arcsec} & \colhead {zeropoint}}
\startdata
F110W &  7 & 0.53 & 14--20 & 1.05--1.50 & 22.141 (008) \\
F160W & 10 & 0.75 & 20--30 & 1.50--2.25 & 21.617 (006) \\
F205W & 14 & 1.05 & 30--40 & 2.25--3.00 & 21.831 (008) \\
\enddata
\end{deluxetable}

\begin{deluxetable}{lrr}
\tablecolumns{3}
\tablewidth{0pt}
\tablenum{3}
\tablecaption{Artificial star tests - Results}
\tablehead{\multicolumn{1}{l}{Field} & \colhead{Offset} &  
\colhead{Log (N/} \\ \colhead{} & \colhead{(mag)} & \colhead{sq pix)}}
\startdata
\ic1613        & 0.007$\pm$0.005 & -2.93 \\
\ic4182        & 0.013$\pm$0.006 & -2.28 \\
M31            & 0.009$\pm$0.007 & -2.19 \\
M81            & 0.085$\pm$0.043 & -1.49 \\
M101-Inner (o) & 0.069$\pm$0.033 & -1.55 \\
M101-Inner (d) & 0.030$\pm$0.033 & -2.16 \\
M101-Outer (o) & 0.026$\pm$0.019 & -2.37 \\
M101-Outer (d) & 0.022$\pm$0.028 & -2.97 \\
\ng925         & 0.084$\pm$0.032 & -1.54 \\
\ng1365        & 0.093$\pm$0.043 & -1.49 \\
\ng2090        & 0.024$\pm$0.029 & -2.05 \\
\ng3198        & 0.073$\pm$0.017 & -1.56 \\
\ng3621        & 0.111$\pm$0.034 & -1.55 \\
\ng4496A       & 0.077$\pm$0.039 & -1.60 \\
\ng4536        & 0.083$\pm$0.031 & -1.57 \\\hline
\multicolumn{3}{l}{{\sc Note:}(d): drizzled; (o): original}
\enddata
\end{deluxetable}

\clearpage

\begin{deluxetable}{lllccc}
\tablenum{4}
\tablecolumns{6}
\tablewidth{0pt}
\tabletypesize{\small}
\baselineskip=8pt
\tablecaption{IC 1613 Secondary standards}
\tablehead{\colhead{Star} & \colhead{R.A.} & \colhead{Dec.} &
\multicolumn{2}{c}{HST} & \colhead{LCO} \\
\colhead{} & \multicolumn{2}{c}{(2000.0)} & \colhead{J} & \colhead{H} &
\colhead{H}}
\startdata
01 & 01:04:43.101 & +02:05:18.31 & $18.43\pm0.14$ & $17.67\pm0.02$ & $17.67\pm0.06$ \\
02 & 01:04:43.775 & +02:05:21.84 & $19.82\pm0.19$ & $19.14\pm0.08$ & $19.23\pm0.07$ \\
03 & 01:04:43.898 & +02:05:25.48 &       \nd      & $18.85\pm0.05$ & $18.84\pm0.03$ \\
04 & 01:04:43.997 & +02:05:23.95 & $18.05\pm0.05$ & $17.23\pm0.02$ & $17.31\pm0.10$ \\
05 & 01:04:44.095 & +02:05:25.68 &       \nd      & $19.21\pm0.08$ & $19.23\pm0.06$ \\
06 & 01:04:44.330 & +02:05:33.52 &       \nd      & $18.97\pm0.05$ & $18.92\pm0.04$ \\
07 & 01:04:44.492 & +02:05:25.27 &       \nd      & $18.78\pm0.04$ & $18.71\pm0.05$ \\
08 & 01:04:44.599 & +02:05:18.94 &       \nd      & $18.91\pm0.10$ & $19.00\pm0.08$ \\
19 & 01:04:47.897 & +02:05:09.59 & $19.35\pm0.17$ & $18.55\pm0.08$ & $18.56\pm0.04$ \\
10 & 01:04:48.192 & +02:05:08.14 & $19.27\pm0.17$ & $18.53\pm0.04$ & $18.62\pm0.01$ \\
11 & 01:04:48.273 & +02:05:06.92 & $18.29\pm0.10$ & $17.46\pm0.02$ & $17.38\pm0.03$ \\
12 & 01:04:50.808 & +02:04:41.49 &       \nd      & $19.91\pm0.11$ & $19.98\pm0.03$ \\
13 & 01:04:51.006 & +02:04:47.48 & $20.41\pm0.32$ & $19.64\pm0.10$ & $19.59\pm0.01$ \\
14 & 01:04:51.142 & +02:05:28.80 & $19.90\pm0.23$ & $19.37\pm0.07$ & $19.36\pm0.08$ \\
15 & 01:04:51.368 & +02:05:29.27 &       \nd      & $19.74\pm0.07$ & $19.78\pm0.07$ \\
16 & 01:04:51.478 & +02:05:32.47 & $19.75\pm0.23$ & $19.01\pm0.05$ & $18.99\pm0.16$ \\
17 & 01:04:51.553 & +02:05:19.08 &       \nd      & $19.01\pm0.07$ & $18.98\pm0.08$ \\
18 & 01:04:51.728 & +02:05:36.74 & $19.83\pm0.25$ & $19.11\pm0.06$ & $19.05\pm0.09$ \\
19 & 01:04:51.732 & +02:05:21.78 &       \nd      & $19.45\pm0.06$ & $19.55\pm0.07$ \\
\multicolumn{4}{l}{\bf Mean $\Delta H$ (LCO-HST):} & \multicolumn{2}{r}{+0.011$\pm$0.061}\\
\enddata
\end{deluxetable}

\clearpage

\begin{deluxetable}{llrllllllll}
\tablecolumns{11}
\tablewidth{0pt}
\tablenum{5}
\tablecaption{Cepheid magnitudes}
\tablehead{\multicolumn{1}{l}{Field} & \colhead{Var.} & \colhead{Per.} & \colhead{H} & \colhead{J}  & \colhead{K}    &
\colhead{V}  & \colhead{I} & \colhead{B}  & \colhead{R}    & \colhead{Ref.}}
\startdata
\ic1613   & V01    &  5.6 & 19.66 (11) & 20.25 (24) &   \ndw     & 20.79      & 20.14      & 21.36      & 20.36       & [1] \\         
          & V14    &  5.1 & 19.49 (07) & 19.81 (19) &   \ndw     & 20.89      & 20.12      & 21.40      & 20.65       & [1] \\         
          & V34    &  8.5 & 19.01 (06) & 19.54 (14) &   \ndw     & 20.74      & 20.03      & 21.41      & 20.50       & [1] \\         
          & V37    & 12.4 & 18.60 (04) & 18.88 (08) &   \ndw     & 20.27      & 19.42      & 21.15      & 19.86       & [1] \\\hline   
\ic4182   & C3-V12 & 36.3 & 20.51 (05) &   \ndw     &   \ndw     & 22.36      & 21.57      &   \ndw     &   \ndw      & [2] \\         
          & C4-V11 & 42.0 & 20.44 (11) &   \ndw     &   \ndw     & 22.33      & 21.40      &   \ndw     &   \ndw      & [2] \\\hline   
M31-F1    & H17    & 18.8 & 18.01 (06) & 19.19 (09) & 17.89 (04) & 19.80 (10) & 19.00 (10) & 20.40 (10) & 19.30 (10)  & [3] \\         
          & V120   & 44.9 & 16.83 (05) & 17.13 (06) & 16.75 (03) & 19.50 (10) & 18.20 (10) & 20.80 (10) & 18.80 (10)  & [3] \\         
{\me}-F3  & H29    & 19.5 & 17.93 (05) & 18.29 (08) & 17.87 (03) & 20.60 (10) & 19.45 (10) & 21.60 (10) & 20.00 (10)  & [3] \\
          & V404   & 17.4 & 18.13 (07) & 18.63 (12) & 18.11 (06) & 20.80 (10) & 19.60 (10) & 21.80 (10) & 20.20 (10)  & [3] \\
          & V427   & 11.3 & 18.88 (06) & 19.18 (11) & 18.81 (04) & 21.00 (10) & 20.05 (10) & 21.80 (10) & 20.50 (10)  & [3] \\
          & V423   & 14.4 & 17.90 (07) & 18.74 (12) & 17.82 (04) & 21.00 (10) & 19.65 (10) & 22.00 (10) & 20.30 (10)  & [3] \\
{\me}-F4  & V08    &  9.6 & 18.82 (05) & 19.24 (07) & 18.80 (06) & 20.40 (10) & 19.70 (10) & 21.10 (10) & 20.10 (10)  & [3] \\         
          & V09    &  8.5 & 19.42 (08) & 20.19 (22) & 19.33 (06) & 20.60 (10) & 20.00 (10) & 21.30 (10) & 20.30 (10)  & [3] \\\hline   
M81       & C06    & 40.8 & 20.33 (08) &   \ndw     &   \ndw     & 22.26      & 21.36      &   \ndw     &   \ndw      & [4] \\         
          & C07    & 27.2 & 20.77 (09) &   \ndw     &   \ndw     & 22.60      & 21.69      &   \ndw     &   \ndw      & [4] \\         
          & C10    & 12.8 & 21.69 (08) &   \ndw     &   \ndw     & 22.91      & 22.29      &   \ndw     &   \ndw      & [4] \\         
          & C11    & 47.2 & 20.05 (10) &   \ndw     &   \ndw     & 22.46      & 21.30      &   \ndw     &   \ndw      & [4] \\         
          & C13    & 18.6 & 21.73 (08) &   \ndw     &   \ndw     & 23.56      & 22.75      &   \ndw     &   \ndw      & [4] \\         
          & C15    & 11.2 & 21.99 (12) &   \ndw     &   \ndw     & 23.84      & 22.96      &   \ndw     &   \ndw      & [4] \\\hline   
M101-     & C051   & 13.0 & 22.78 (15) & 23.39 (32) &   \ndw     & 24.89 (03) & 23.96 (05) &   \ndw     &   \ndw      & [5] \\         
Inner     & C161   & 23.9 & 21.78 (05) & 22.49 (20) &   \ndw     & 24.36 (02) & 23.33 (03) &   \ndw     &   \ndw      & [5] \\         
          & C172   & 15.4 & 22.93 (09) & 23.46 (42) &   \ndw     & 24.92 (02) & 23.84 (04) &   \ndw     &   \ndw      & [5] \\         
          & C186   & 27.8 & 21.87 (05) & 22.55 (22) &   \ndw     & 25.14 (03) & 23.55 (04) &   \ndw     &   \ndw      & [5] \\         
          & C192   & 73.8 & 20.89 (03) & 21.19 (12) &   \ndw     & 23.00 (02) & 21.89 (02) &   \ndw     &   \ndw      & [5] \\         
          & C194   & 44.8 & 21.47 (04) & 22.27 (16) &   \ndw     & 24.26 (02) & 22.93 (03) &   \ndw     &   \ndw      & [5] \\         
          & C205   & 26.1 & 22.02 (05) & 22.33 (24) &   \ndw     & 23.56 (01) & 22.80 (03) &   \ndw     &   \ndw      & [5] \\         
          & C212   & 29.4 & 21.91 (05) & 22.18 (19) &   \ndw     & 24.09 (01) & 23.04 (01) &   \ndw     &   \ndw      & [5] \\\hline   
M101-     & C01    & 58.5 & 21.14 (04) & 21.55 (09) &   \ndw     & 23.83 (07) & 22.42 (09) &   \ndw     &   \ndw      & [6] \\         
Outer     & C06    & 45.8 & 21.37 (06) & 21.81 (13) &   \ndw     & 23.47 (07) & 22.62 (13) &   \ndw     &   \ndw      & [6] \\         
          & C07    & 43.0 & 22.04 (07) & 22.42 (18) &   \ndw     & 23.75 (08) & 22.84 (08) &   \ndw     &   \ndw      & [6] \\         
          & C08    & 41.0 & 21.69 (07) & 22.45 (41) &   \ndw     & 23.88 (08) & 23.00 (09) &   \ndw     &   \ndw      & [6] \\         
          & C10    & 37.6 & 22.18 (05) & 22.65 (12) &   \ndw     & 24.01 (08) & 22.94 (16) &   \ndw     &   \ndw      & [6] \\         
          & C20    & 42.5 & 21.91 (06) & 22.61 (16) &   \ndw     & 24.11 (07) & 22.93 (08) &   \ndw     &   \ndw      & [6] \\         
          & C24    & 23.5 & 22.46 (09) & 22.72 (15) &   \ndw     & 24.25 (09) & 23.55 (09) &   \ndw     &   \ndw      & [6] \\         
          & C26    & 17.7 & 23.03 (19) & 23.33 (18) &   \ndw     & 24.66 (09) & 23.82 (09) &   \ndw     &   \ndw      & [6] \\
\enddata
\end{deluxetable}

\begin{deluxetable}{llrllllllll}
\tablecolumns{11}
\tablewidth{0pt}
\tablenum{5}
\tablecaption{Cepheid magnitudes -- continued}
\tablehead{\multicolumn{1}{l}{Field} & \colhead{Var.} & \colhead{Per.} & \colhead{H} & \colhead{J}  & \colhead{K}    &
\colhead{V}  & \colhead{I} & \colhead{B}  & \colhead{R}    & \colhead{Ref.}}
\startdata
\ng925    & C06    & 43.2 & 22.48 (09) &   \ndw     &   \ndw     & 24.55 (10) & 23.47 (10) &   \ndw     &   \ndw      & [7] \\         
          & C08    & 37.3 & 22.23 (07) &   \ndw     &   \ndw     & 24.65 (10) & 23.63 (10) &   \ndw     &   \ndw      & [7] \\         
          & C09    & 35.1 & 22.36 (09) &   \ndw     &   \ndw     & 24.79 (10) & 23.74 (10) &   \ndw     &   \ndw      & [7] \\         
          & C13    & 30.4 & 22.60 (15) &   \ndw     &   \ndw     & 24.64 (10) & 23.55 (10) &   \ndw     &   \ndw      & [7] \\         
          & C17    & 28.5 & 23.14 (10) &   \ndw     &   \ndw     & 25.14 (10) & 23.94 (10) &   \ndw     &   \ndw      & [7] \\         
          & C24    & 25.3 & 22.52 (09) &   \ndw     &   \ndw     & 25.11 (10) & 23.94 (10) &   \ndw     &   \ndw      & [7] \\         
          & C26    & 23.7 & 22.83 (10) &   \ndw     &   \ndw     & 25.02 (10) & 23.98 (10) &   \ndw     &   \ndw      & [7] \\         
          & C33    & 21.5 & 22.86 (11) &   \ndw     &   \ndw     & 24.54 (10) & 23.92 (10) &   \ndw     &   \ndw      & [7] \\         
          & C41    & 18.3 & 23.47 (26) &   \ndw     &   \ndw     & 25.55 (10) & 24.59 (10) &   \ndw     &   \ndw      & [7] \\         
          & C50    & 16.4 & 23.45 (14) &   \ndw     &   \ndw     & 25.01 (10) & 24.41 (10) &   \ndw     &   \ndw      & [7] \\
\ng1365   & C01    & 60.0 & 23.40 (24) &   \ndw     &   \ndw     & 25.66 (07) & 24.40 (07) &   \ndw     &   \ndw      & [8] \\         
          & C04    & 55.0 & 22.77 (11) &   \ndw     &   \ndw     & 25.52 (06) & 24.29 (06) &   \ndw     &   \ndw      & [8] \\         
          & C06    & 47.0 & 23.84 (31) &   \ndw     &   \ndw     & 26.08 (09) & 25.19 (09) &   \ndw     &   \ndw      & [8] \\\hline   
\ng2090   & C03    & 48.8 & 23.25 (16) &   \ndw     &   \ndw     & 25.07      & 24.07      &   \ndw     &   \ndw      & [9] \\          
          & C18    & 23.7 & 23.70 (21) &   \ndw     &   \ndw     & 25.61      & 24.69      &   \ndw     &   \ndw      & [9] \\          
          & C21    & 18.5 & 23.81 (20) &   \ndw     &   \ndw     & 25.60      & 24.78      &   \ndw     &   \ndw      & [9] \\          
          & C23    & 17.3 & 23.82 (22) &   \ndw     &   \ndw     & 25.51      & 24.84      &   \ndw     &   \ndw      & [9] \\\hline    
\ng3198   & C03    & 26.4 & 23.03 (12) &   \ndw     &   \ndw     & 25.32 (12) & 24.44 (09) &   \ndw     &   \ndw      & [10]\\         
          & C06    & 18.7 & 23.71 (22) &   \ndw     &   \ndw     & 25.82 (09) & 25.05 (12) &   \ndw     &   \ndw      & [10]\\         
          & C10    & 45.1 & 22.68 (10) &   \ndw     &   \ndw     & 24.97 (10) & 23.98 (11) &   \ndw     &   \ndw      & [10]\\\hline   
\ng3621   & C16    & 31.2 & 21.99 (12) &   \ndw     &   \ndw     & 24.50 (03) & 23.28 (04) &   \ndw     &   \ndw      & [11]\\         
          & C17    & 28.3 & 21.86 (07) &   \ndw     &   \ndw     & 23.89 (03) & 22.84 (04) &   \ndw     &   \ndw      & [11]\\         
          & C32    & 23.5 & 22.00 (14) &   \ndw     &   \ndw     & 24.11 (04) & 23.26 (08) &   \ndw     &   \ndw      & [11]\\         
          & C35    & 22.8 & 22.78 (10) &   \ndw     &   \ndw     & 24.77 (03) & 23.60 (05) &   \ndw     &   \ndw      & [11]\\         
          & C66    & 11.9 & 22.87 (10) &   \ndw     &   \ndw     & 25.57 (06) & 24.23 (08) &   \ndw     &   \ndw      & [11]\\\hline
\ng4496A  & C44    & 36.1 & 23.89 (15) &   \ndw     &   \ndw     & 25.65 (03) & 24.61 (05) &   \ndw     &   \ndw      & [12]\\         
          & C46    & 25.1 & 24.36 (31) &   \ndw     &   \ndw     & 25.72 (02) & 24.97 (05) &   \ndw     &   \ndw      & [12]\\         
          & C47    & 38.5 & 23.54 (11) &   \ndw     &   \ndw     & 25.50 (02) & 24.67 (05) &   \ndw     &   \ndw      & [12]\\         
          & C50    & 46.2 & 23.13 (07) &   \ndw     &   \ndw     & 25.41 (02) & 24.28 (03) &   \ndw     &   \ndw      & [12]\\         
          & C59    & 22.2 & 24.20 (19) &   \ndw     &   \ndw     & 25.98 (03) & 24.95 (08) &   \ndw     &   \ndw      & [12]\\         
          & C60    & 33.8 & 23.56 (16) &   \ndw     &   \ndw     & 25.04 (02) & 24.37 (04) &   \ndw     &   \ndw      & [12]\\         
          & C67    & 39.3 & 23.68 (12) &   \ndw     &   \ndw     & 26.02 (03) & 24.82 (05) &   \ndw     &   \ndw      & [12]\\\hline   
\ng4536   & C2-V4  & 28.7 & 24.18 (18) &   \ndw     &   \ndw     & 25.81 (18) & 25.06 (15) &   \ndw     &   \ndw      & [13]\\         
          & C2-V9  & 58.0 & 23.50 (11) &   \ndw     &   \ndw     & 25.39 (12) & 24.45 (11) &   \ndw     &   \ndw      & [13]\\         
\enddata
\tablerefs{[1]: \citet{fr88}; [2]: \citet{sah94}; [3]: Madore (priv. comm.);
[4]:~\citet{fr94} [5]: \citet{ste98}; [6]: \citet{kel96}; [7]: \citet{sil96};
[8]: \citet{sil99}; [9]: \citet{phe98}; [10]: \citet{kel99};
[11]:~\citet{raw97}; [12]: \citet{gib00}; [13]: \citet{sah96}}
\end{deluxetable}

\clearpage

\begin{deluxetable}{lcccccccc}
\tablecolumns{9}
\tablewidth{0pt}
\tablenum{6}
\tablecaption{Observed $H$, $I$ and $V$ distance moduli}
\tablehead{\multicolumn{1}{l}{Field} & \multicolumn{4}{c}{This work} & \multicolumn{4}{c}{Published} \\
\colhead{} & \colhead{$\mu_H$} & \colhead{$\mu_I$} & \colhead{$\mu_V$} & 
\colhead{$N$} & \colhead{$\mu_I$} & \colhead{$\mu_V$} & \colhead{$N$} & \colhead{Ref.}}
\startdata
\ic1613    & 24.43 (08) & 24.44 (13) & 24.53 (13) &  4 & 24.39 (09) & 24.50 (09) &  9 & [1] \\
\ic4182    & 28.05 (07) & 28.14 (01) & 28.20 (07) &  2 & 28.33 (06) & 28.37 (07) & 27 & [2] \\
M31        & 24.54 (08) & 24.94 (09) & 25.23 (17) &  9 & 24.76 (05) & 25.01 (07) & 37 & [2] \\
M81        & 27.91 (08) & 28.02 (13) & 28.15 (17) &  6 & 28.03 (07) & 28.22 (09) & 17 & [2] \\
M101-Inner & 29.04 (08) & 29.37 (10) & 29.71 (18) &  8 & 29.31 (06) & 29.49 (08) & 61 & [3] \\
M101-Outer & 29.45 (08) & 29.58 (09) & 29.77 (09) &  8 & 29.33 (05) & 29.46 (07) & 25 & [2] \\
\ng0925    & 29.84 (08) & 30.08 (05) & 30.30 (10) & 10 & 30.12 (03) & 30.33 (04) & 72 & [2] \\
\ng1365    & 31.36 (27) & 31.69 (21) & 31.99 (11) &  3 & 31.49 (04) & 31.69 (05) & 47 & [2] \\
\ng2090    & 30.57 (20) & 30.66 (13) & 30.75 (18) &  4 & 30.54 (04) & 30.71 (05) & 30 & [2] \\
\ng3198    & 30.25 (12) & 30.72 (07) & 30.83 (09) &  3 & 30.89 (04) & 31.04 (05) & 36 & [2] \\
\ng3621    & 29.09 (15) & 29.38 (10) & 29.75 (16) &  5 & 29.61 (05) & 29.97 (07) & 59 & [2] \\
\ng4496A   & 31.12 (07) & 31.12 (09) & 31.29 (14) &  7 & 31.00 (03) & 31.14 (03) & 94 & [2] \\
\ng4536    & 31.47 (15) & 31.46 (15) & 31.51 (21) &  2 & 31.06 (04) & 31.24 (04) & 35 & [2] \\
\enddata
\tablerefs{[1]: based on data from \citet{fr88}; [2]: derived by \citet{fr01b}; [3]: derived by \citet{ste98}.}
\end{deluxetable}

\begin{deluxetable}{lccr}
\tablecolumns{4}
\tablewidth{0pt}
\tablenum{7}
\tablecaption{Observed $J$ and $K$ distance moduli}
\tablehead{\multicolumn{1}{l}{Field} & \colhead{Filter\ \ } & \colhead{$\mu$} & \multicolumn{1}{r}
{$N$}}
\startdata
\ic1613    & J & 24.53 (12) & 4 \\
M31        & J & 24.77 (13) & 9 \\
M31        & K & 24.55 (08) & 9 \\
M101-Inner & J & 29.19 (08) & 8 \\
M101-Outer \ \ \ \ & J & 29.53 (03) &  \ \ \  \ \ 8 \\
\enddata
\end{deluxetable}

\clearpage

\begin{deluxetable}{lcr}
\tablecolumns{3}
\tablewidth{0pt}
\tablenum{8}
\tablecaption{Mean color excesses}
\tablehead{\multicolumn{1}{l}{Field} & \colhead{$\evi$} & \colhead{$\evh$}}
\startdata
\ic1613           & 0.09 (03) & 0.10 (09) \\
\ic4182           & 0.06 (05) & 0.15 (00) \\
M31               & 0.29 (08) & 0.69 (20) \\
M81               & 0.13 (05) & 0.24 (11) \\
M101-Inner        & 0.36 (08) & 0.67 (17) \\
M101-Outer        & 0.21 (07) & 0.31 (10) \\
\ng0925           & 0.24 (06) & 0.47 (09) \\
\ng1365           & 0.32 (09) & 0.63 (13) \\
\ng2090           & 0.11 (05) & 0.18 (04) \\
\ng3198$^\dagger$ & 0.13 (03) & 0.58 (03) \\
\ng3621           & 0.40 (08) & 0.66 (15) \\
\ng4496A          & 0.19 (07) & 0.17 (12) \\
\ng4536           & 0.06 (05) & 0.03 (04) \\
\multicolumn{2}{l}{\bf Mean ratio:}          &2.02$\pm$0.22 \\
\multicolumn{2}{l}{\bf Zeropoint: }          &-0.05$\pm$0.05\\
\enddata
\tablecomments{$\dagger$: $4\sigma$ outlier, rejected from fit.}
\end{deluxetable}

\clearpage

\begin{figure}
\plotone{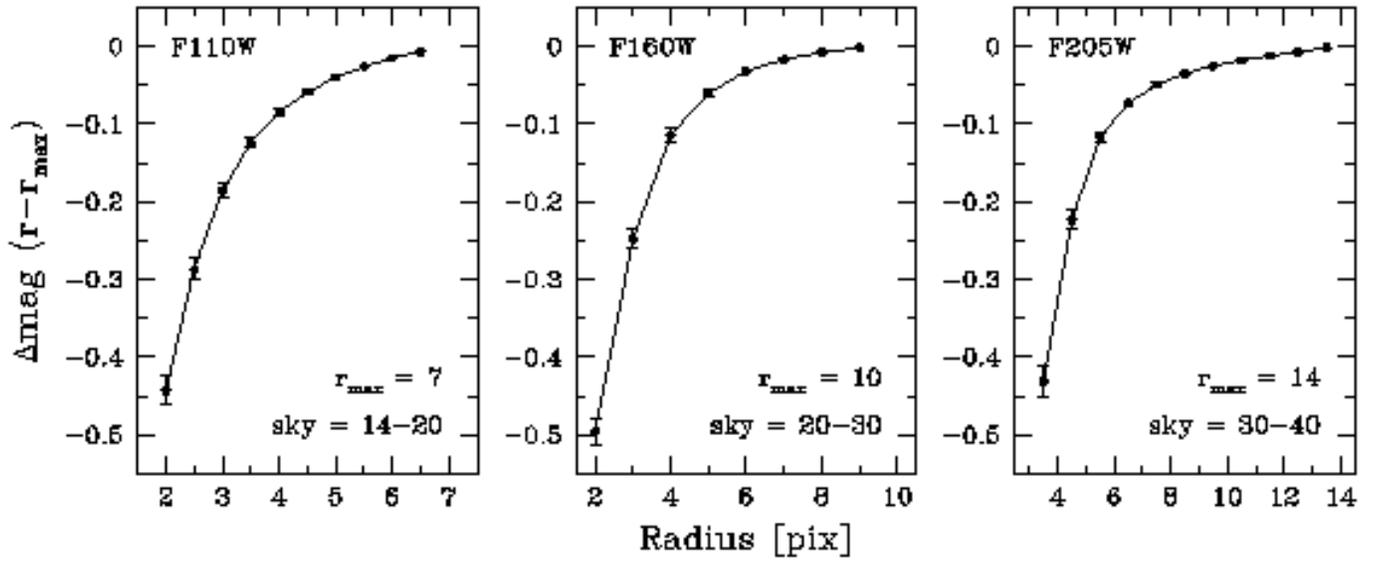}
\caption{Cumulative growth curves for the F110W, F160W and F205W
bands derived by {DAOGROW}. Aperture and inner and outer sky radii are
listed in each panel and in Table 2. One pixel equals $0\farcs 075$.}
\end{figure}

\clearpage

\begin{figure}
\plotone{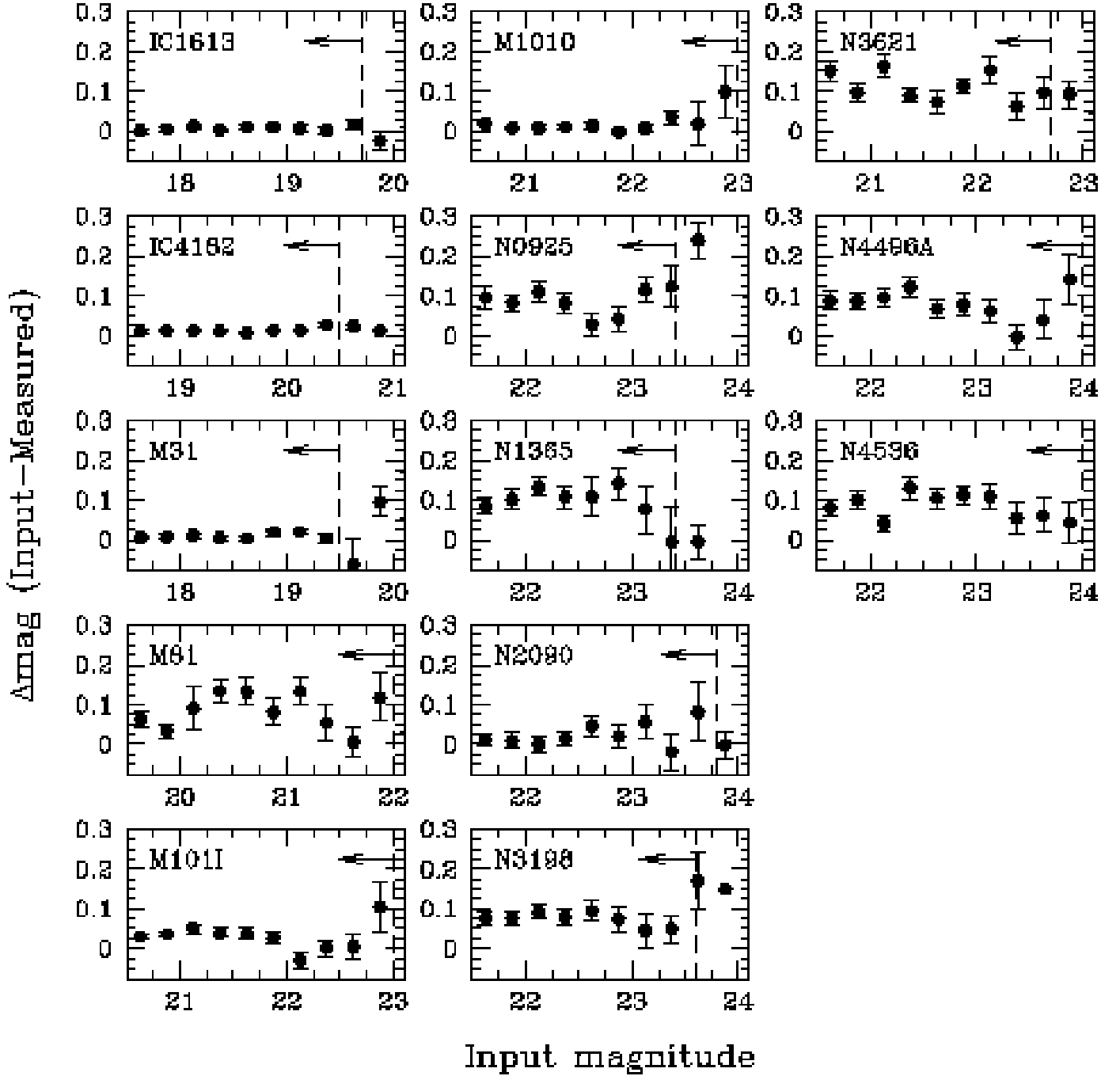}
\caption{(a) Results of the photometric recovery tests described in
\S3.3. Each panel represents one of the field/filter combinations which
yielded useful data. Input magnitudes are plotted on the ordinate, while
the abscissa shows differences between the input and the recovered
magnitudes. Averages over 0.25~mag are plotted using filled circles. The dashed
vertical lines mark the faint end of the Cepheid distribution.}
\end{figure}

\clearpage

\setcounter{figure}{1}

\begin{figure}
\plotone{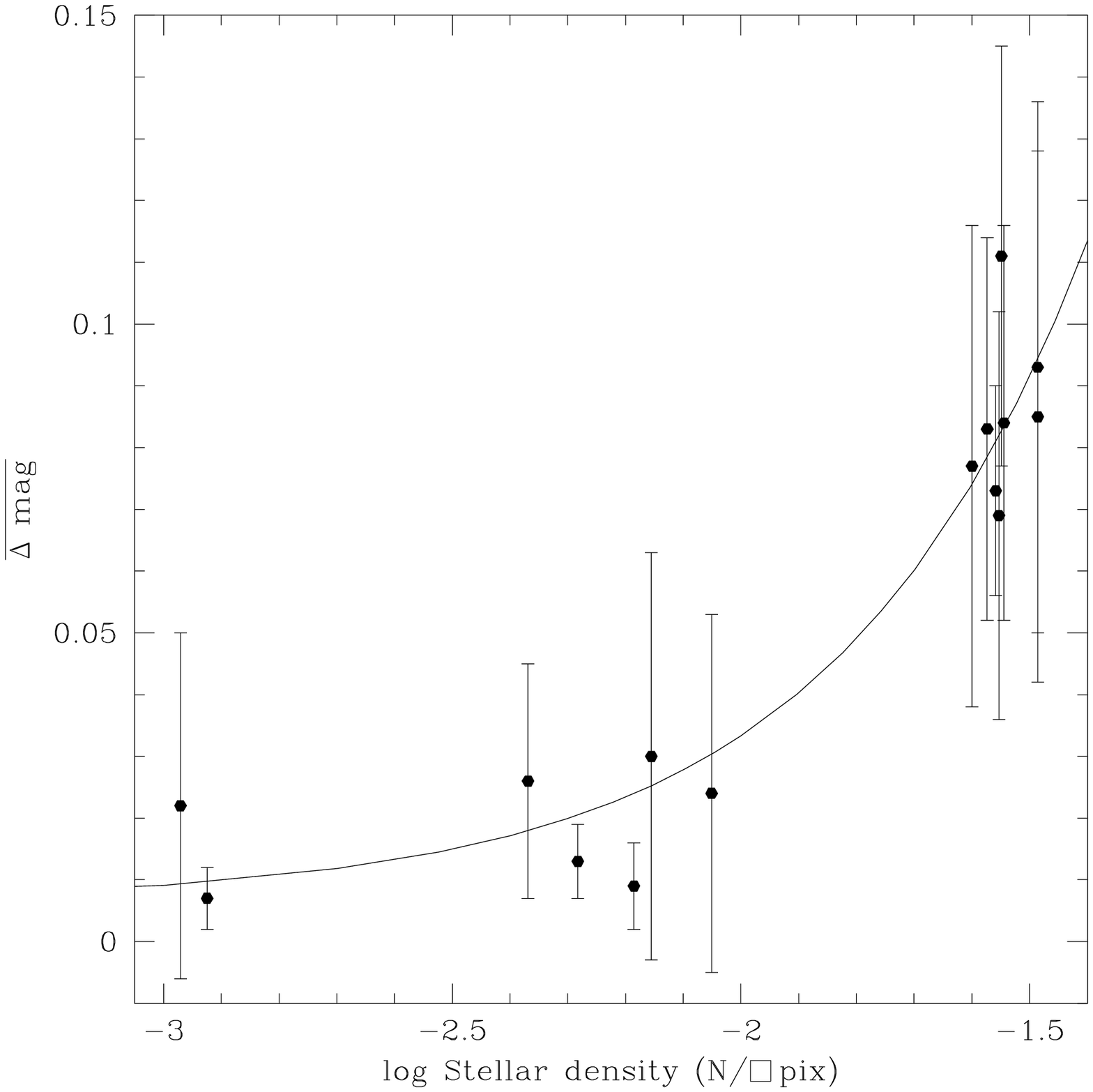}
\caption{(b) Results of the photometric recovery tests described in
\S3.3. Correlation between stellar density and crowding bias; filled
circles represent the data found in Table 3. The solid line is derived
from a linear least-squares fit to the data.}
\end{figure}

\clearpage

\begin{figure}
\plotone{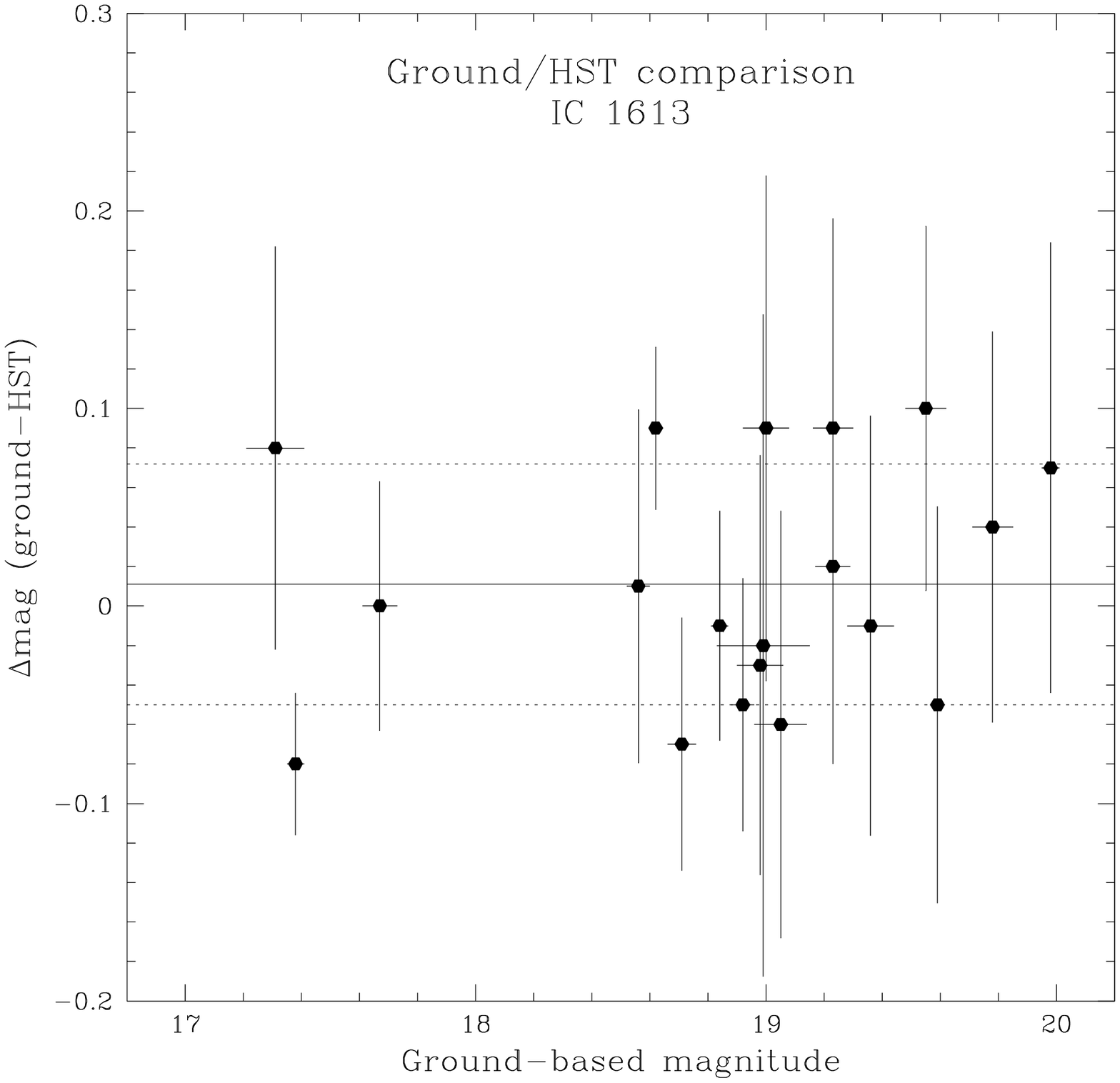}
\caption{Comparison of ground-based and HST $H$-band photometry for
bright, isolated stars in the \ic1613 field listed in Table 4. A mean offset of
$0.011\pm0.061$~mag is indicated by solid and dashed lines.}
\end{figure}

\clearpage

\begin{figure}
\plotone{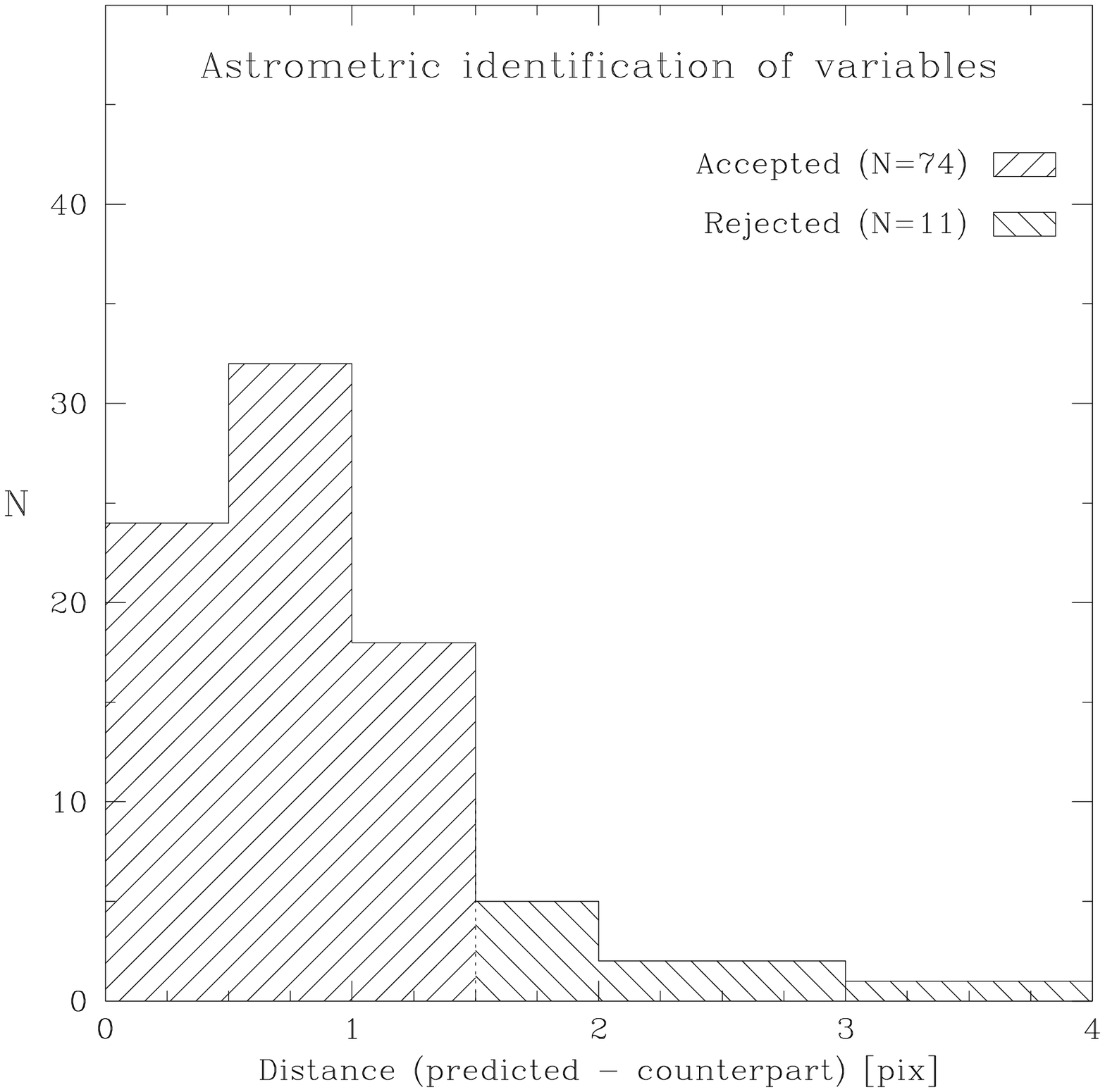}
\caption{Histogram of radial distances between predicted positions
of our variables and the location of the nearest object in the frame. We
rejected 11 candidates whose distance between these two positions was larger
than 1.5 pixels.}
\end{figure}

\clearpage

\begin{figure}
\plotone{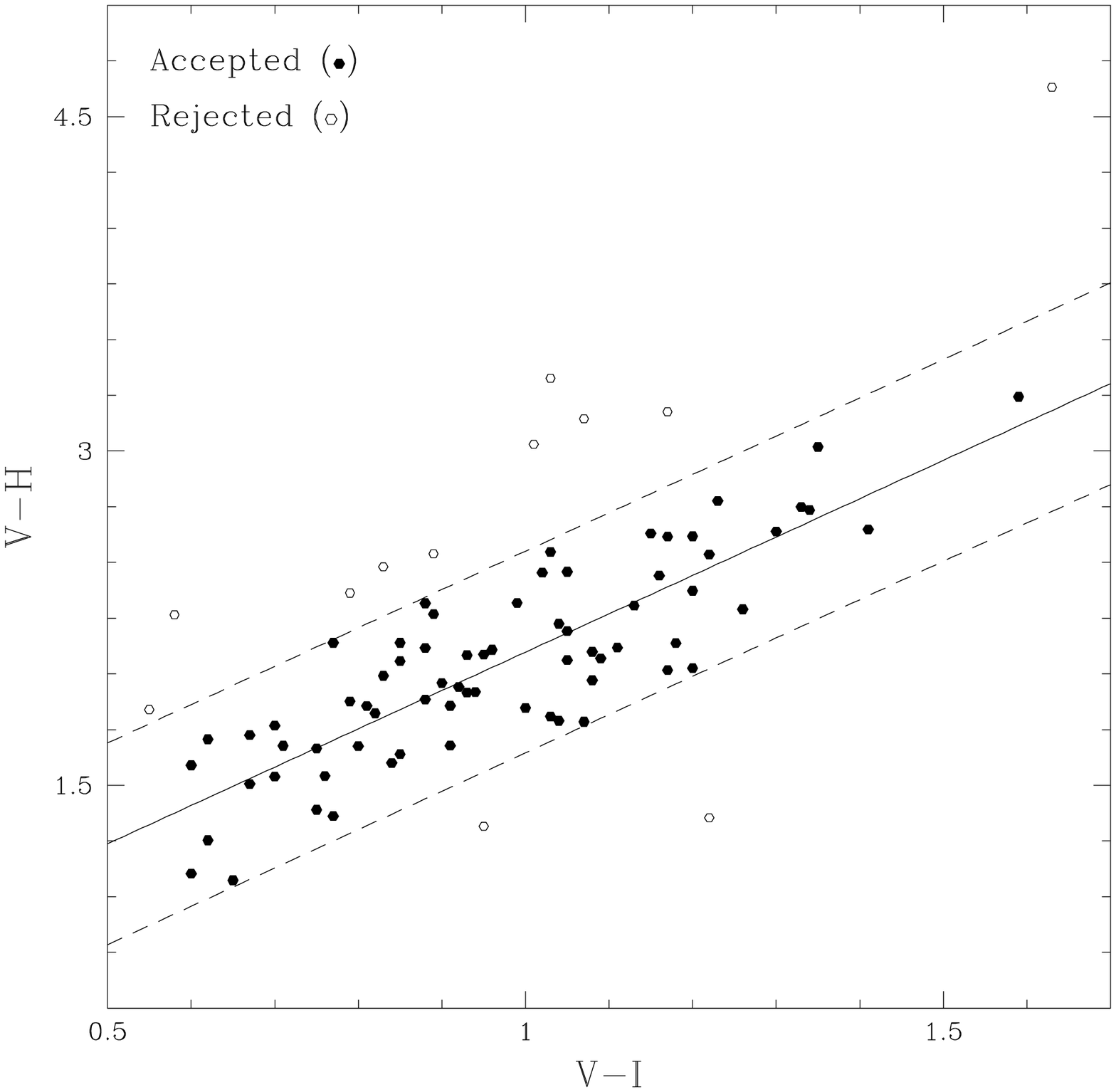}
\caption{(a) Color-color diagram for the 82 candidates which passed the
astrometry test. Filled and open symbols indicate the candidates that passed
and failed this test, respectively. The solid line is a least-squares fit to a
color/color relation with a slope of 1.72 (see text for details).}
\end{figure}

\clearpage

\setcounter{figure}{4}

\begin{figure}
\plotone{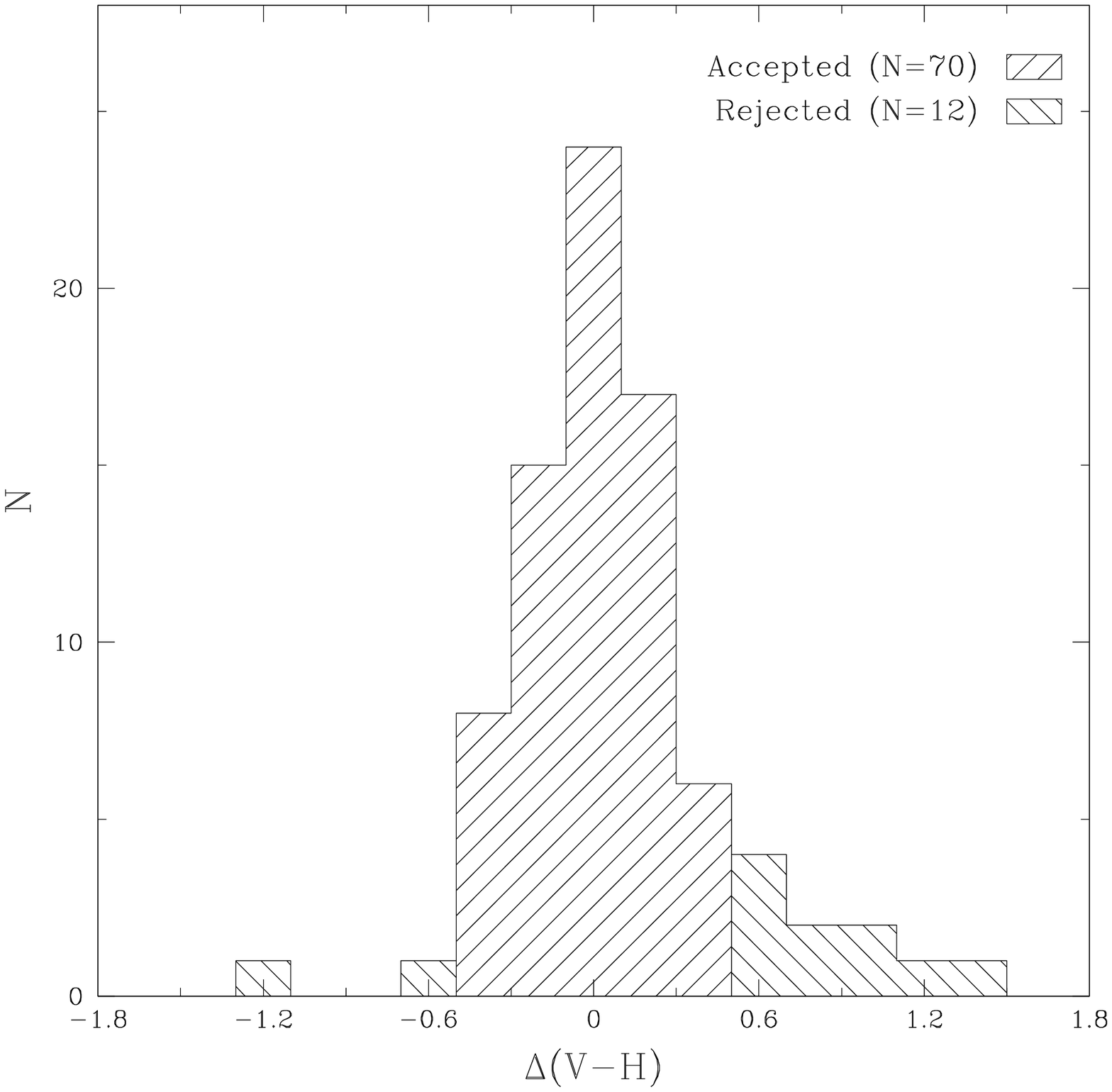}
\caption{(b) Histogram of the deviations from the best-fit line of Fig 5a.
We rejected 12 candidates with deviations larger than twice the {\it rms}.}
\end{figure}

\clearpage

\begin{figure}
\epsscale{0.8}
\plotone{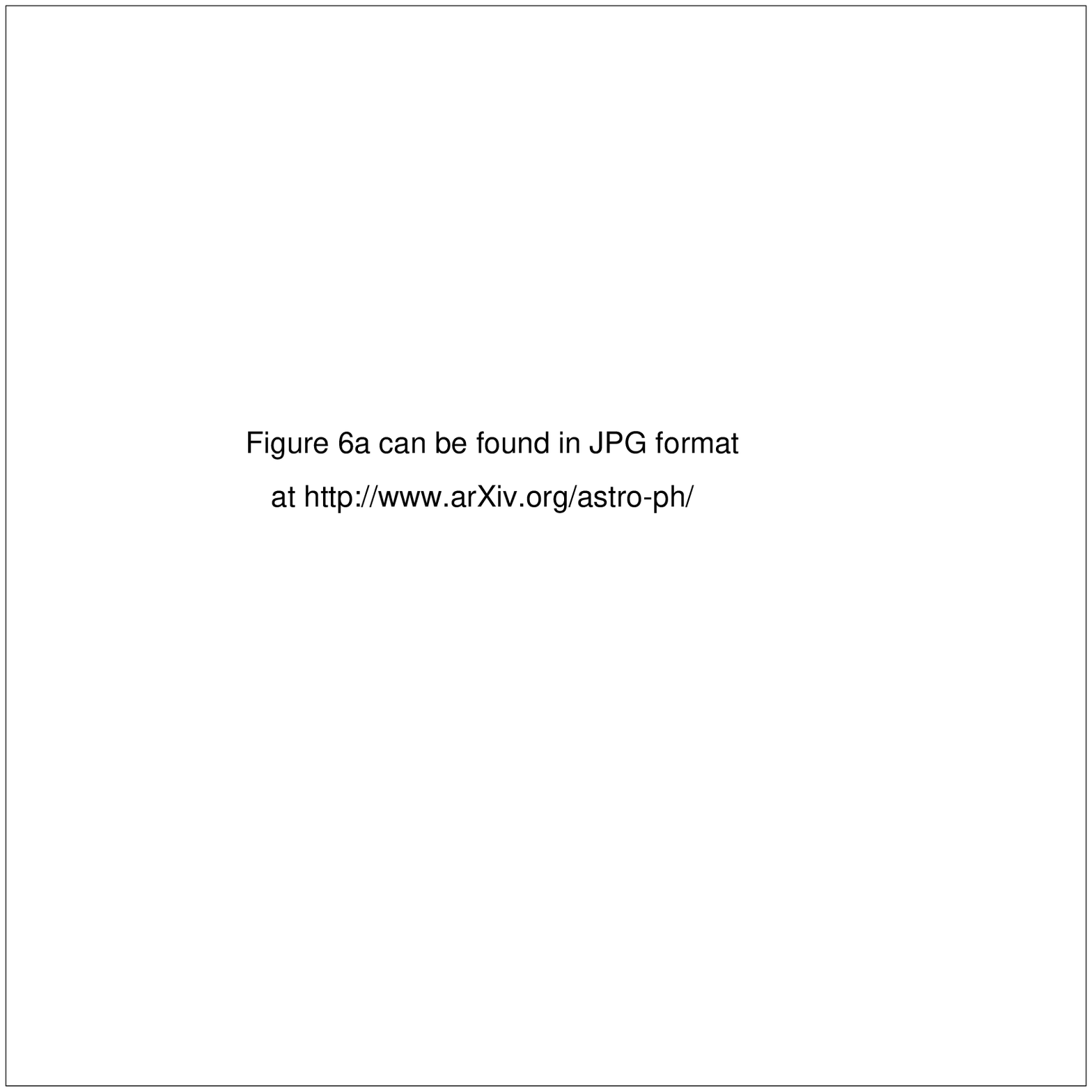}
\caption{(a) Finding charts for all fields used in this work. The images are
full-frame F160W mosaics, spanning $\sim 24\arcsec$. The galaxy name and the
STScI-assigned mosaic identification are displayed in each image. Small circles
indicate the location of the Cepheids used in this work. The variable names
match those of Table 5.}
\end{figure}

\clearpage

\setcounter{figure}{5}

\begin{figure}
\epsscale{0.9}
\plotone{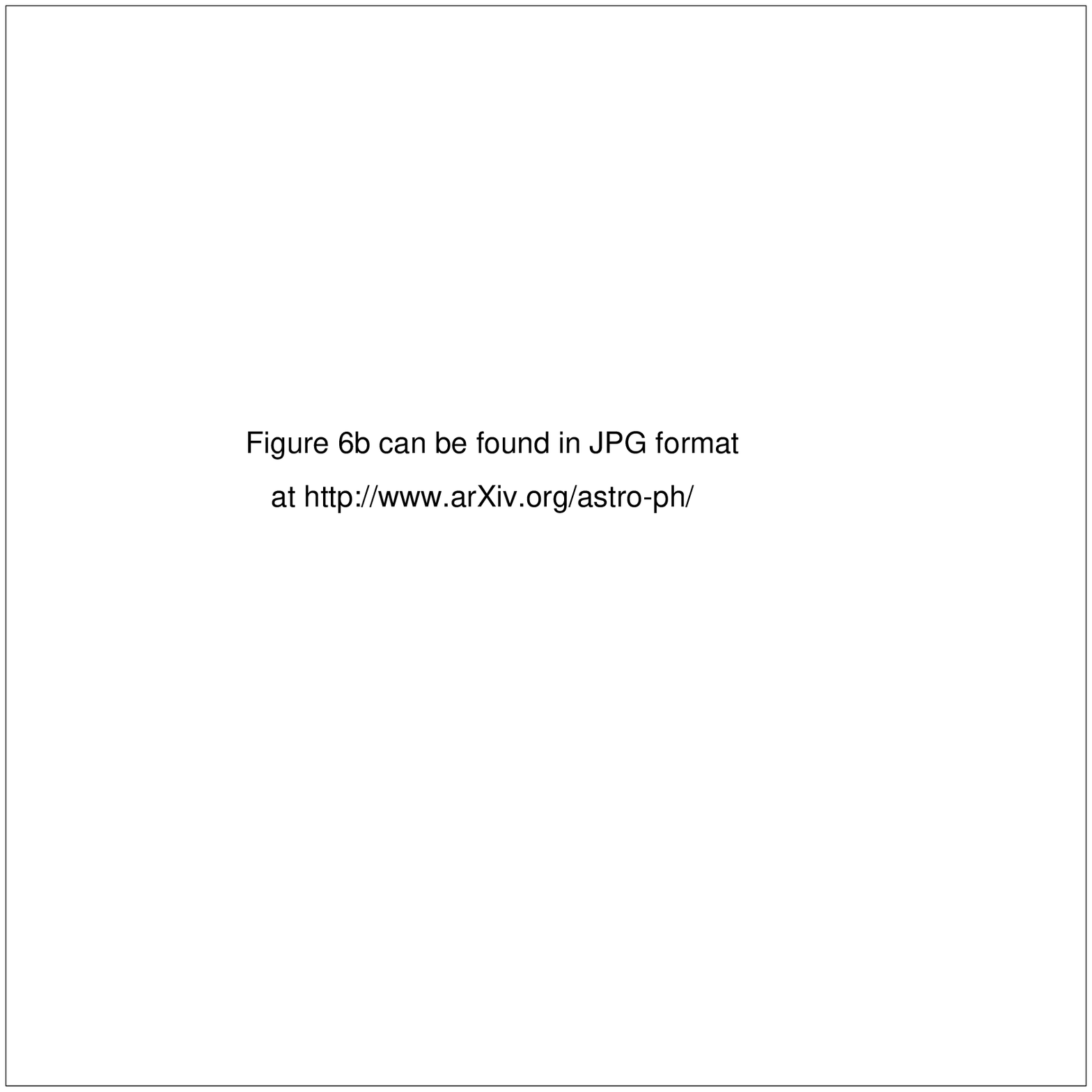}
\caption{(b) (continued)}
\end{figure}

\clearpage

\setcounter{figure}{5}

\begin{figure}
\plotone{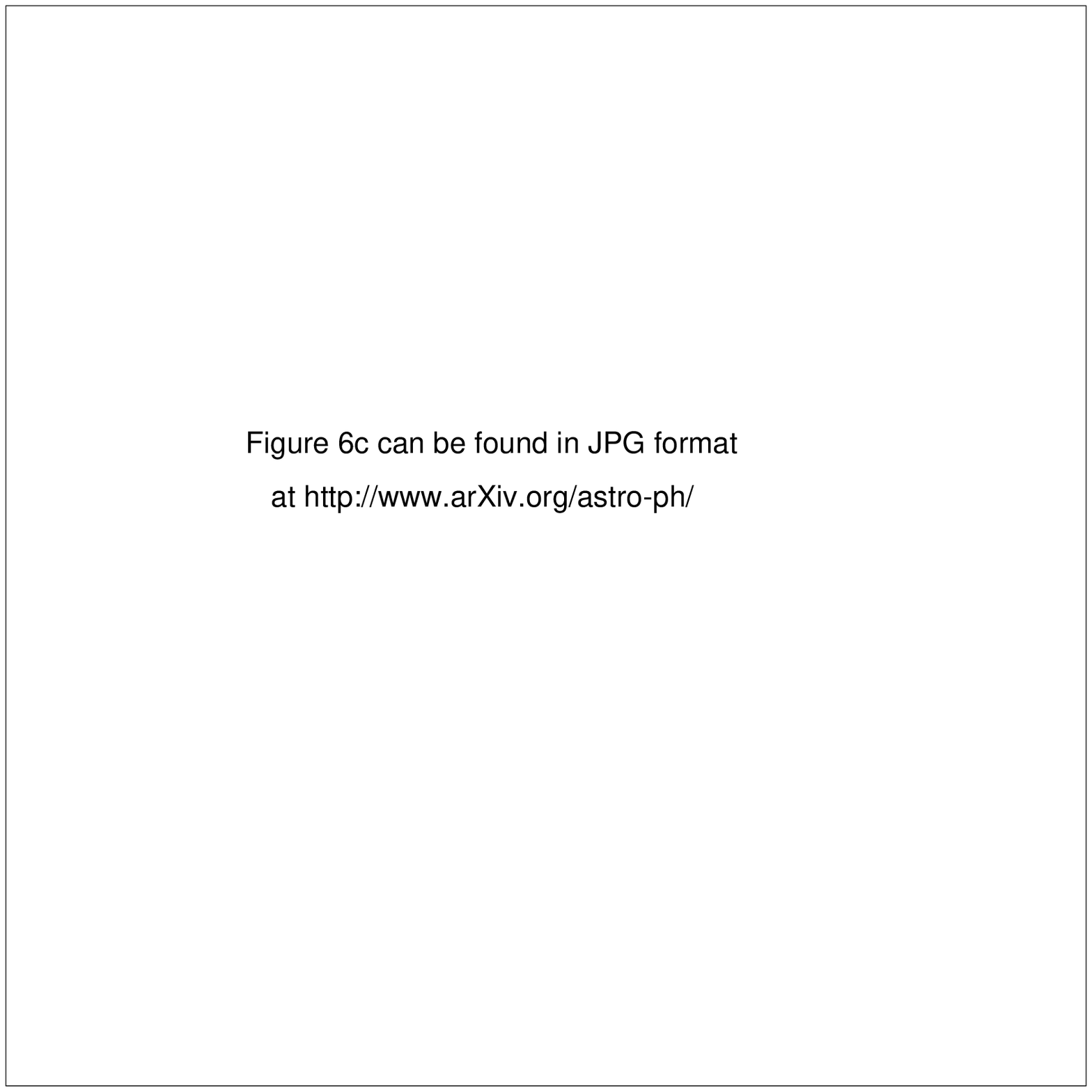}
\caption{(c) (continued)}
\end{figure}

\clearpage

\setcounter{figure}{5}

\begin{figure}
\plotone{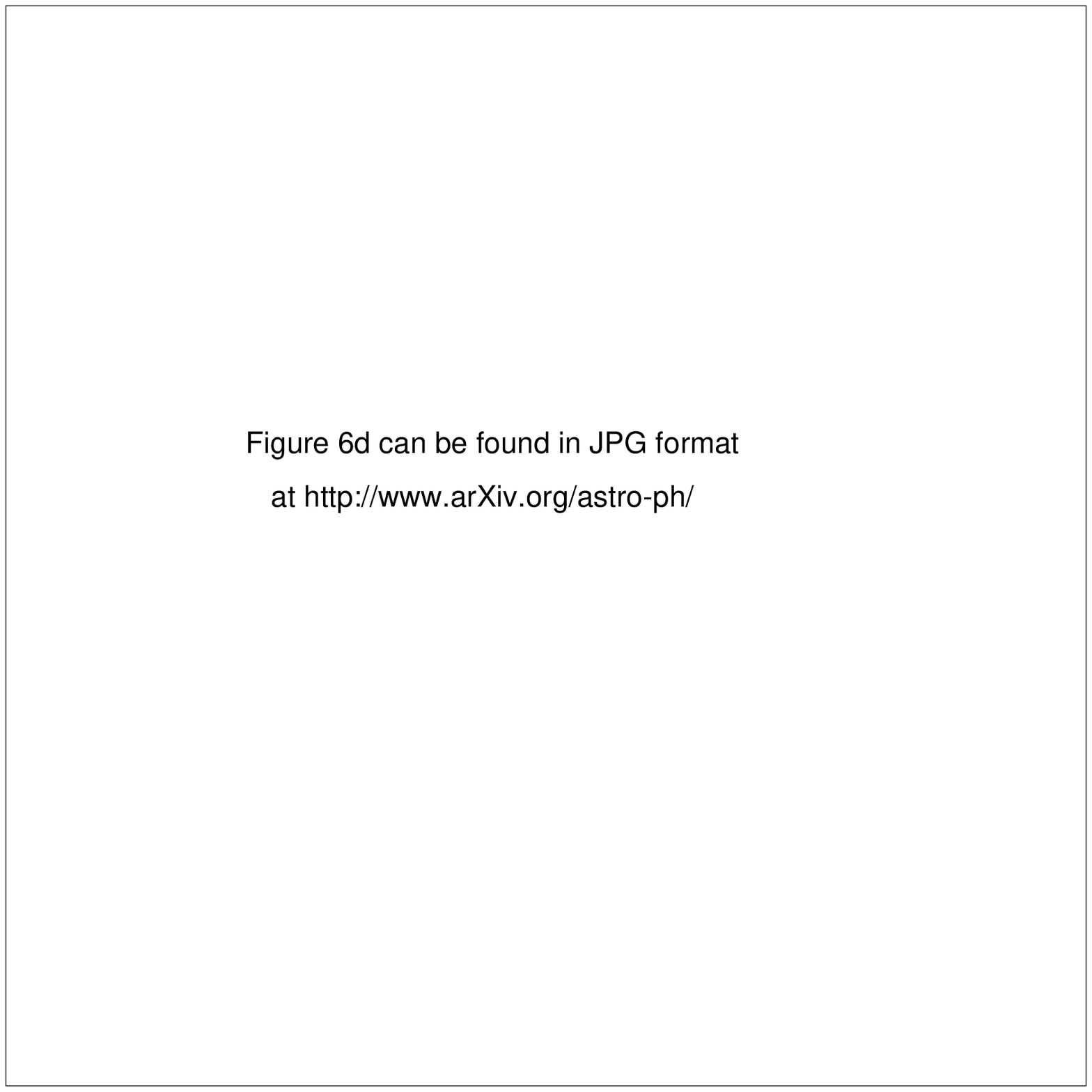}
\caption{(d) (continued)}
\end{figure}

\clearpage

\setcounter{figure}{5}

\begin{figure}
\plotone{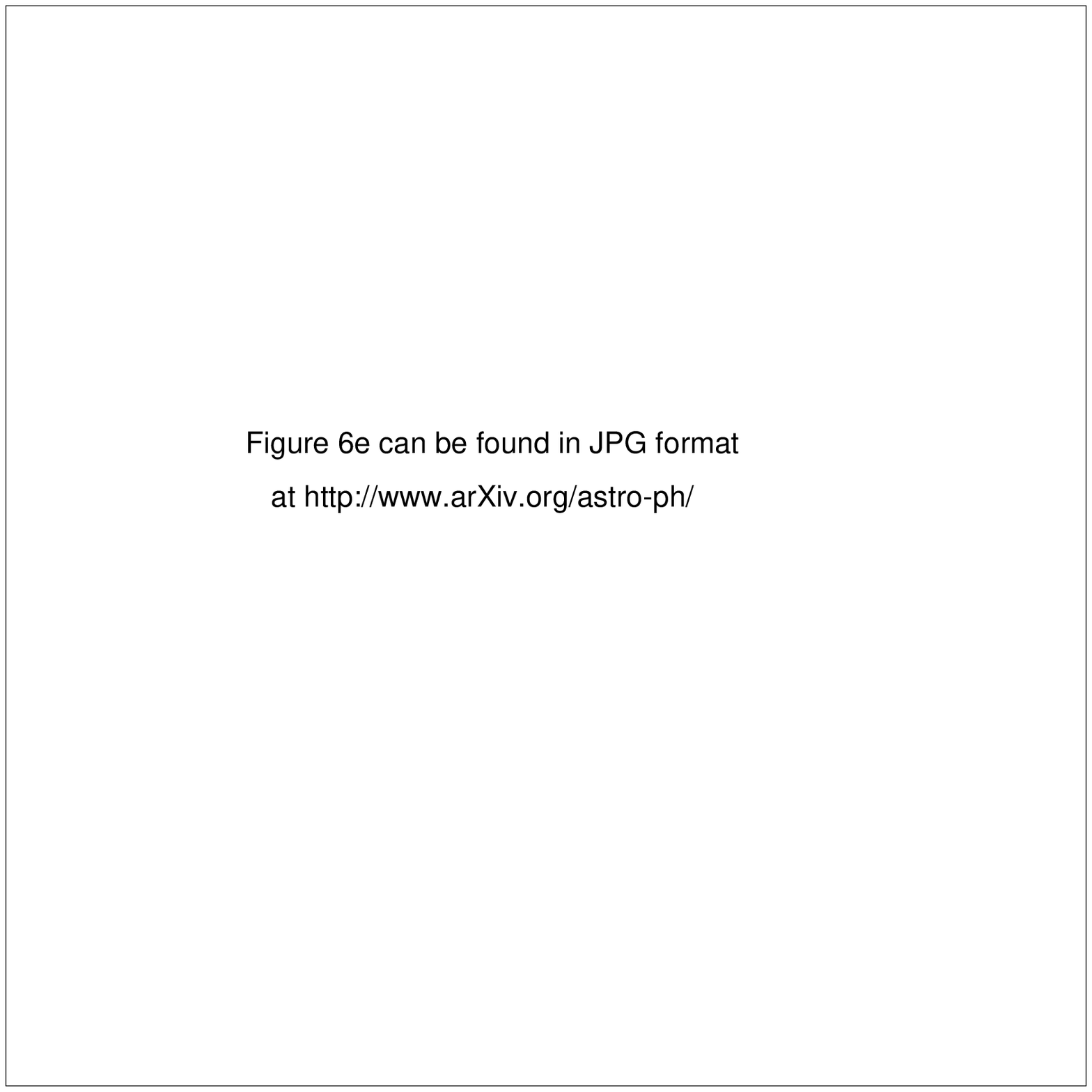}
\caption{(e) (continued)}
\end{figure}

\clearpage

\setcounter{figure}{5}

\begin{figure}
\plotone{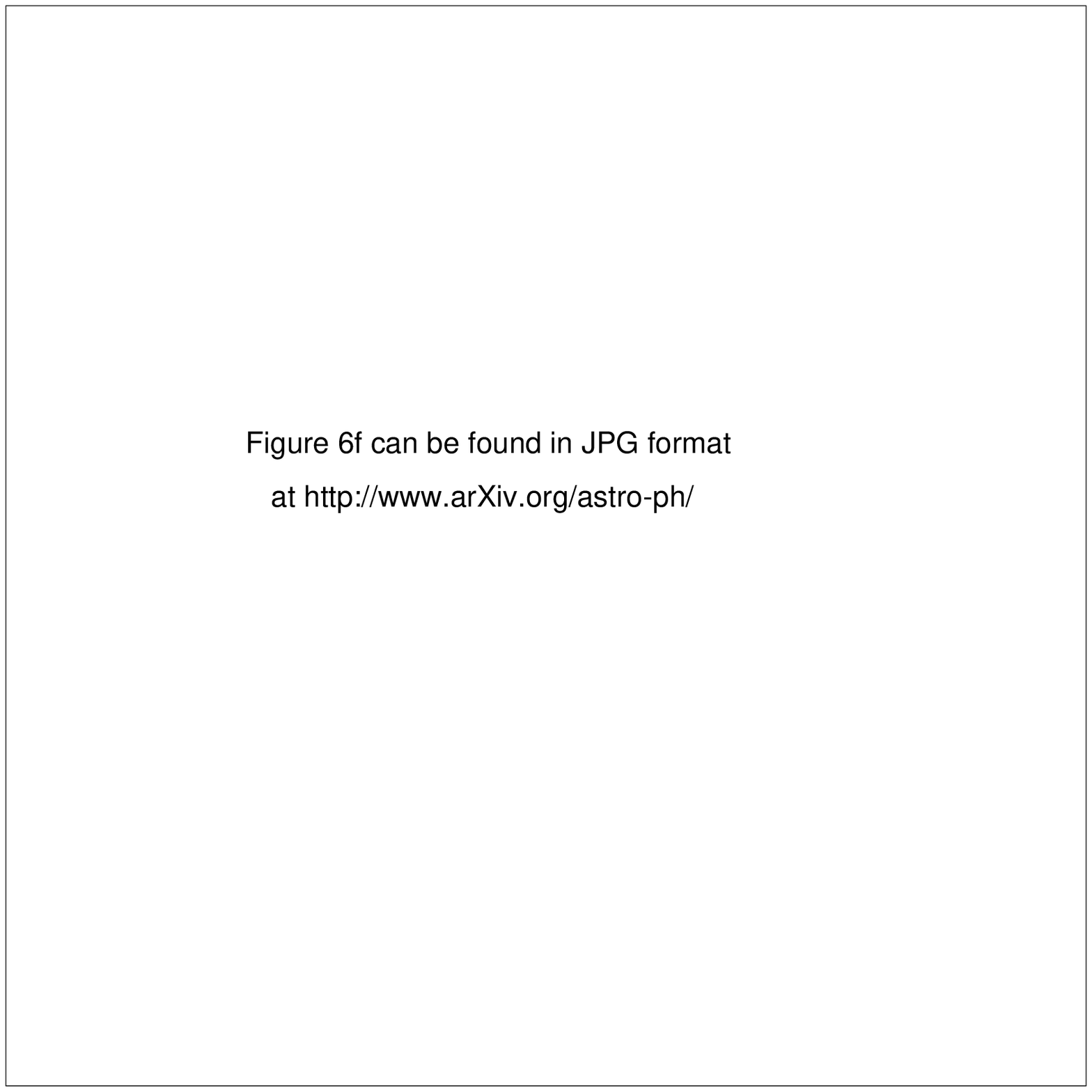}
\caption{(f) (continued)}
\end{figure}

\clearpage

\begin{figure}
\plotone{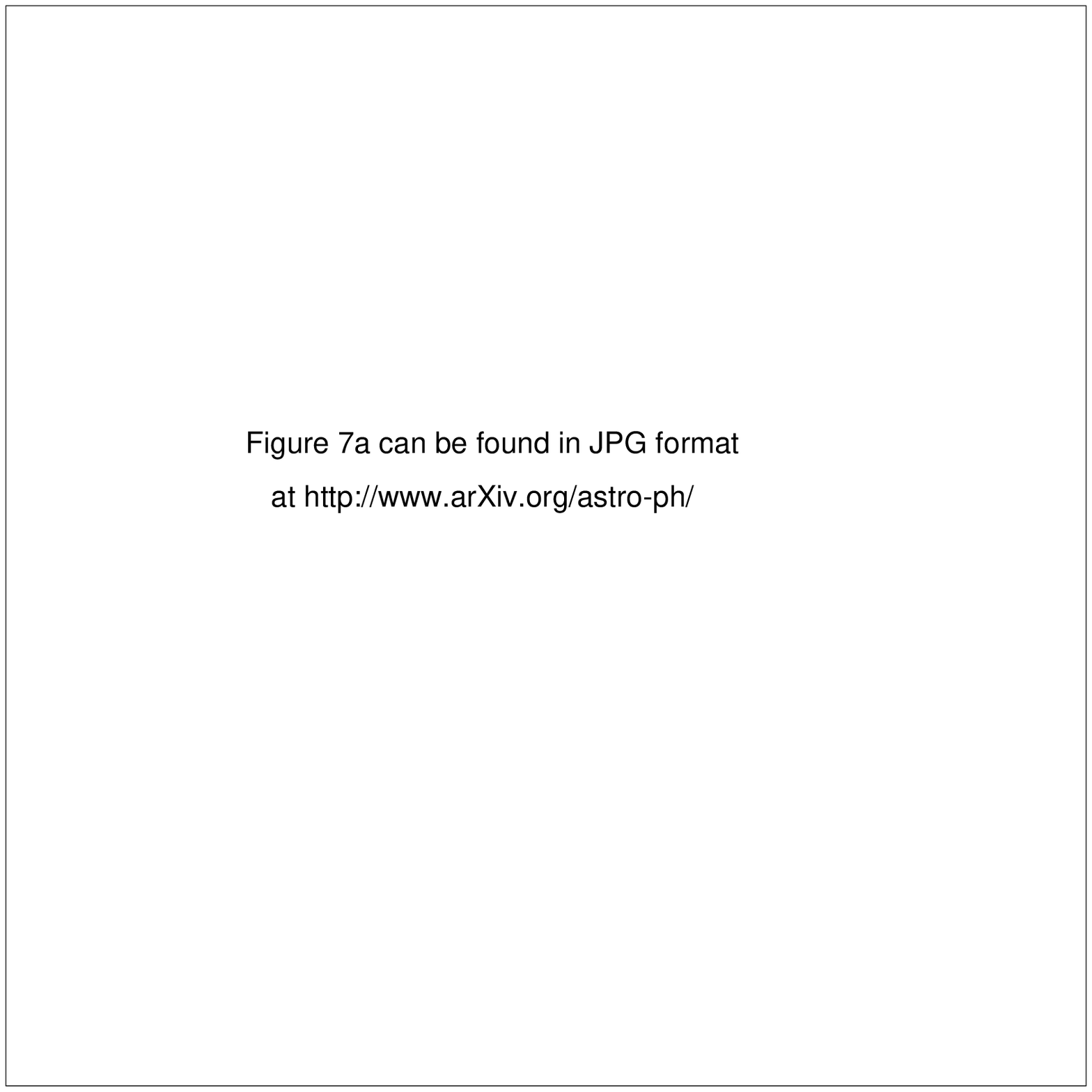}
\caption{(a) Individual finding charts for all Cepheids used in this work.
Each box is 38 pixels or $\sim 3\arcsec$ on a side. Cepheids are marked by a
small circle. The designation of each star is shown in the upper-left corner of
each box.}
\end{figure}

\clearpage

\setcounter{figure}{6}

\begin{figure}
\plotone{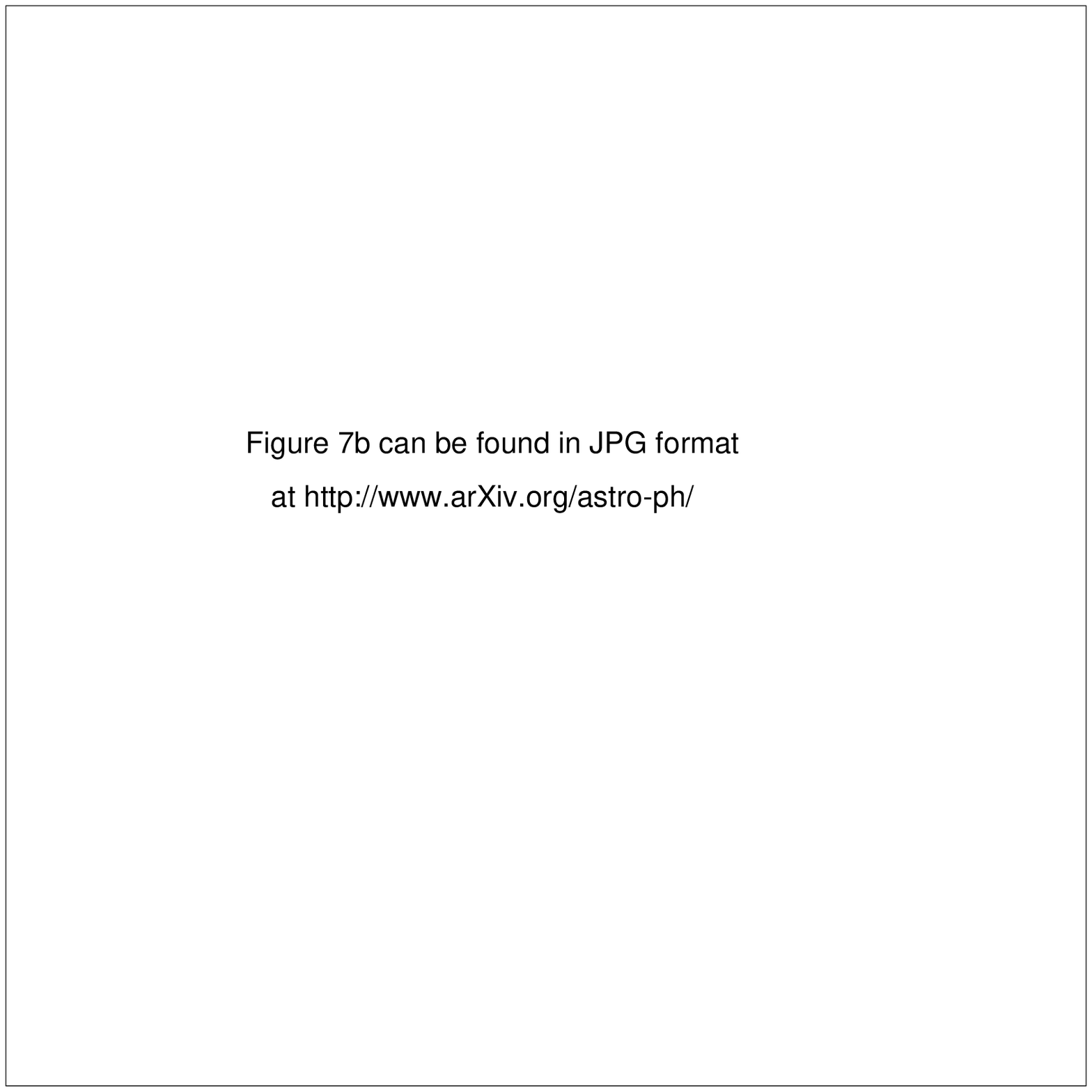}
\caption{(b) (continued)}
\end{figure}

\clearpage

\begin{figure}
\plotone{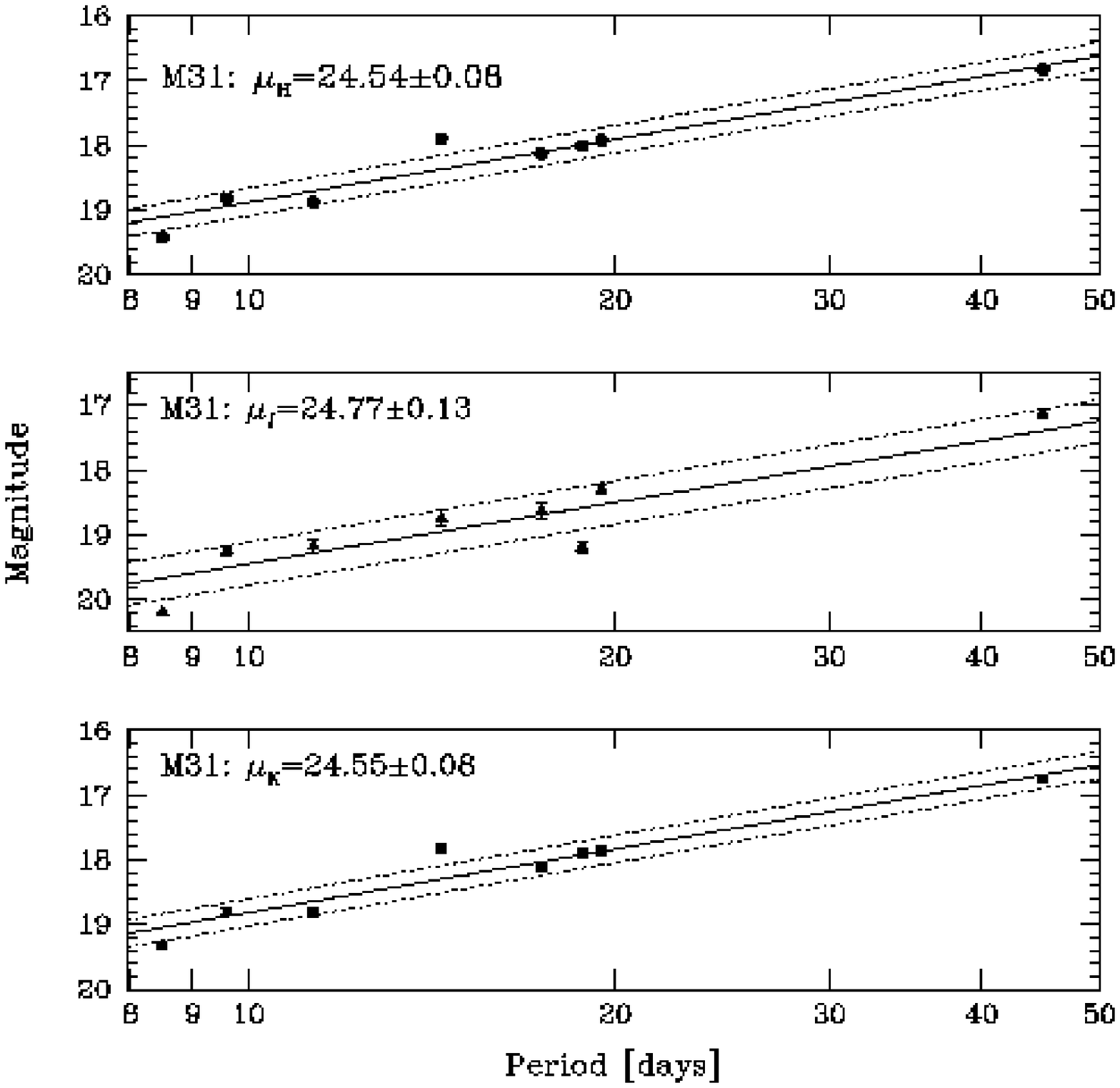}
\caption{(a) Near-infrared Period-Luminosity relations for all field/filter
combinations. Field name, observed distance modulus and its uncertainty
appear in the top-left corner of each panel. Solid lines indicate
fitting results while dashed ones indicate the {\it rms} uncertainties.}
\end{figure}

\clearpage

\setcounter{figure}{7}
\begin{figure}
\plotone{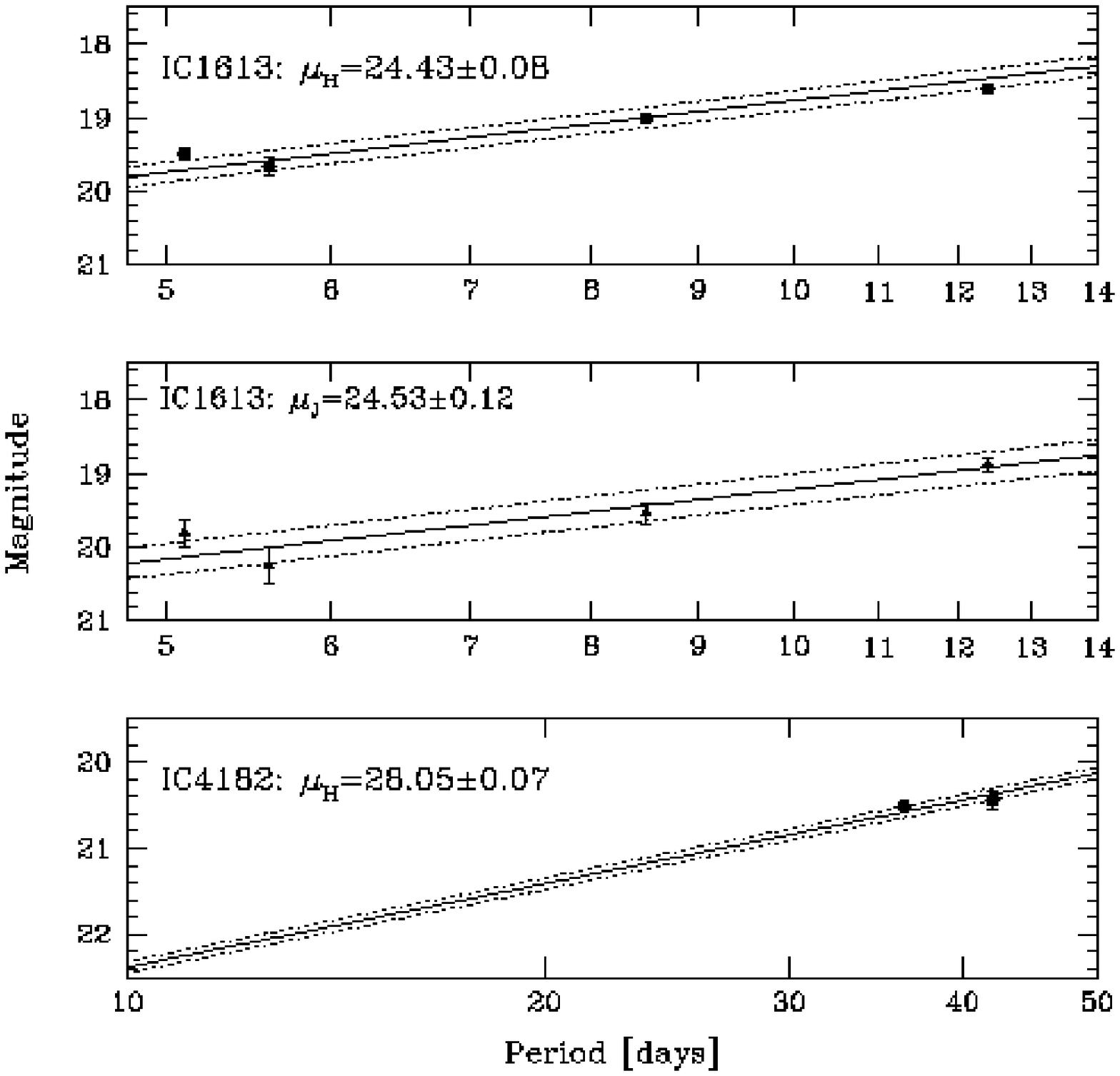}
\caption{(b) (continued)}
\end{figure}

\clearpage

\setcounter{figure}{7}

\begin{figure}
\plotone{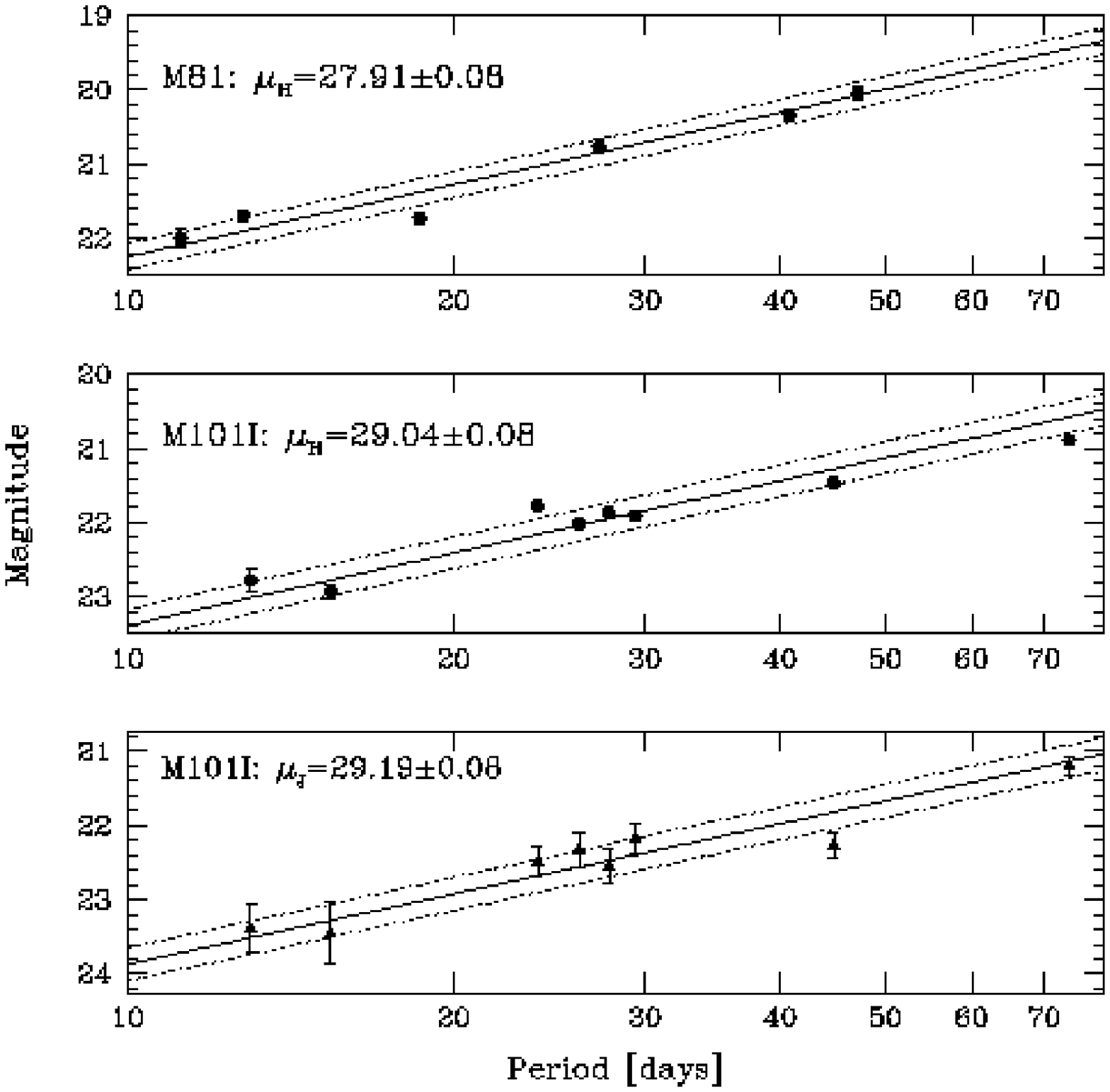}
\caption{(c) (continued)}
\end{figure}

\clearpage

\setcounter{figure}{7}

\begin{figure}
\plotone{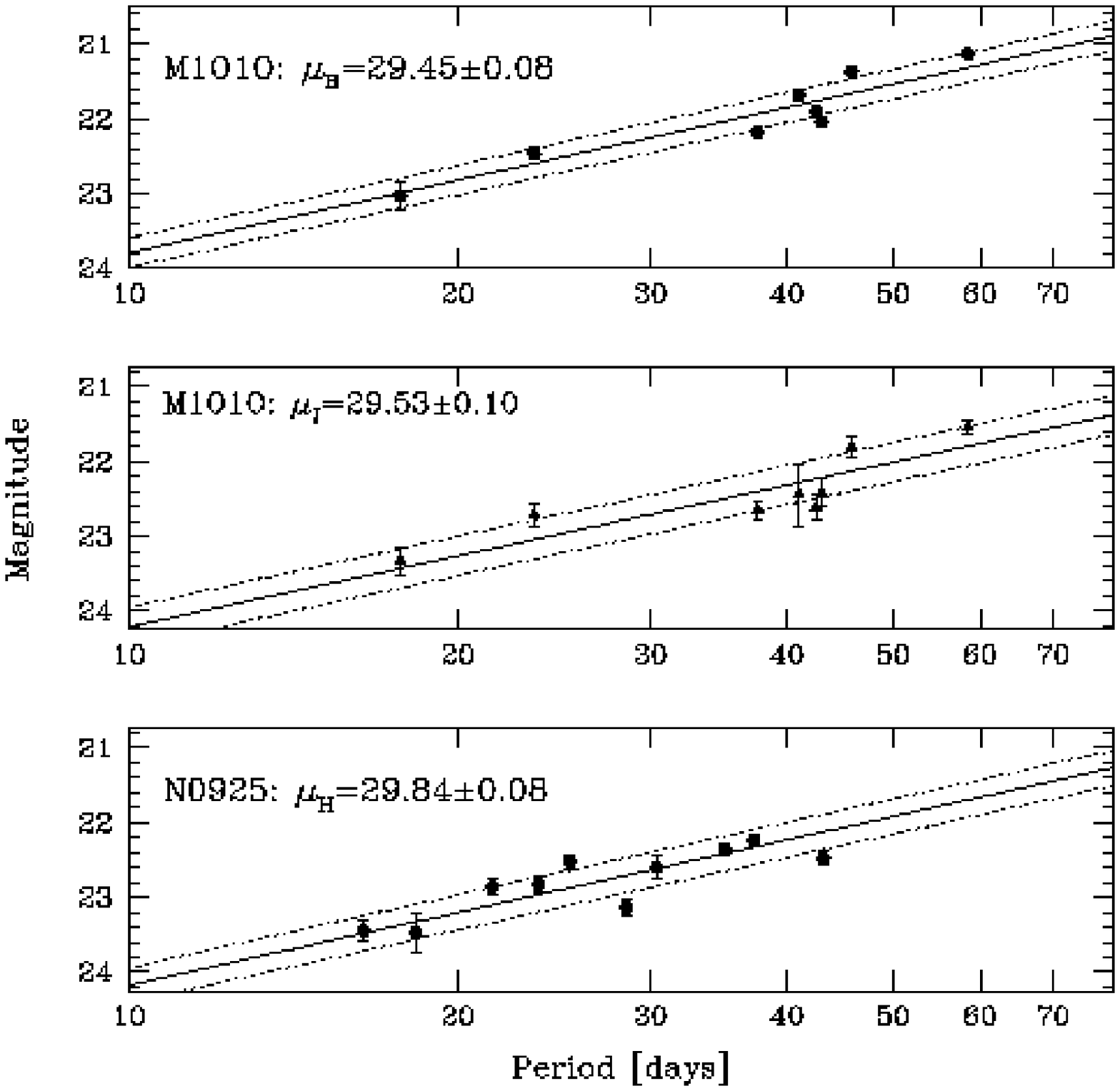}
\caption{(d) (continued)}
\end{figure}

\clearpage

\setcounter{figure}{7}

\begin{figure}
\plotone{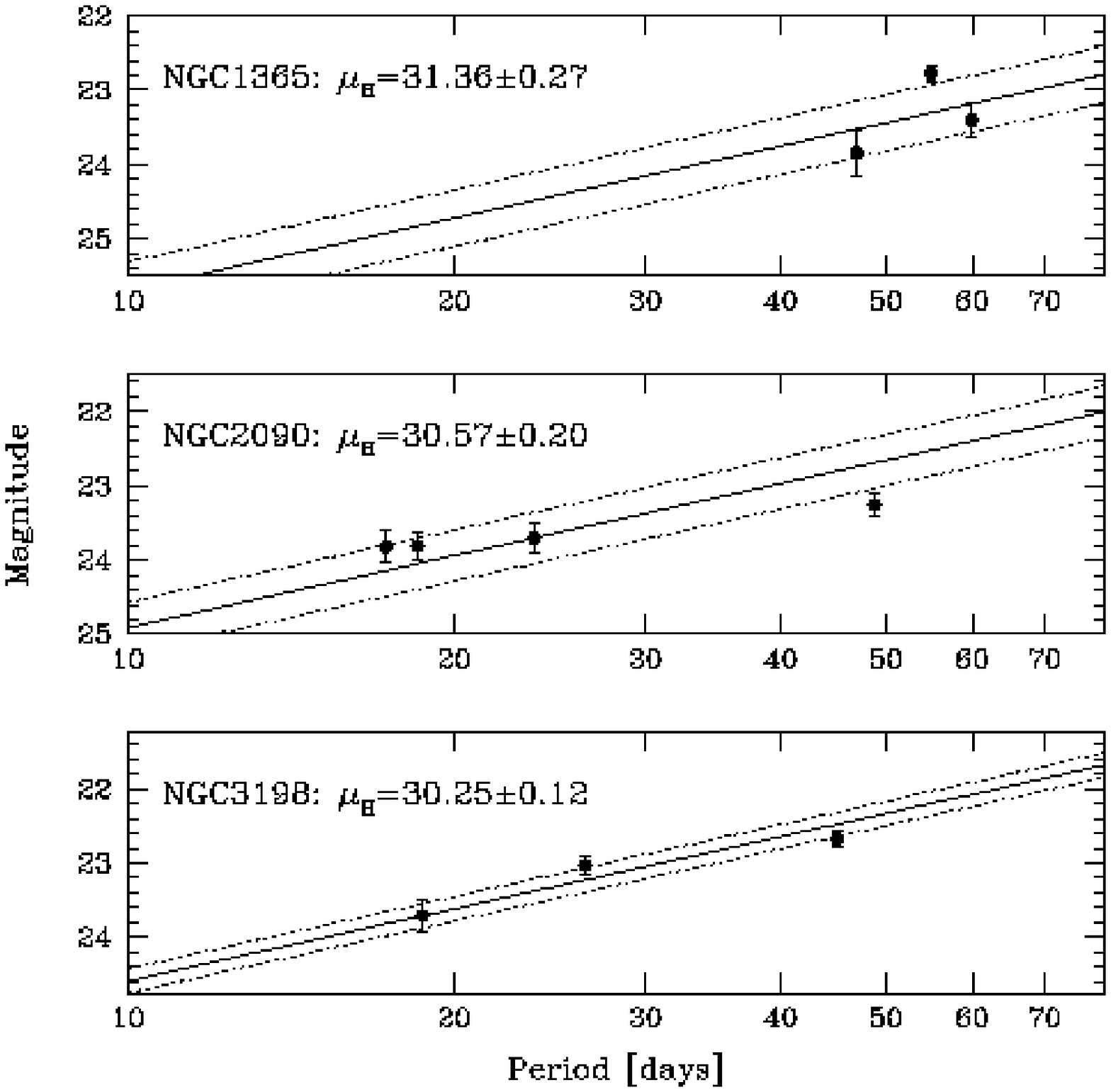}
\caption{(e) (continued)}
\end{figure}

\clearpage

\setcounter{figure}{7}

\begin{figure}
\plotone{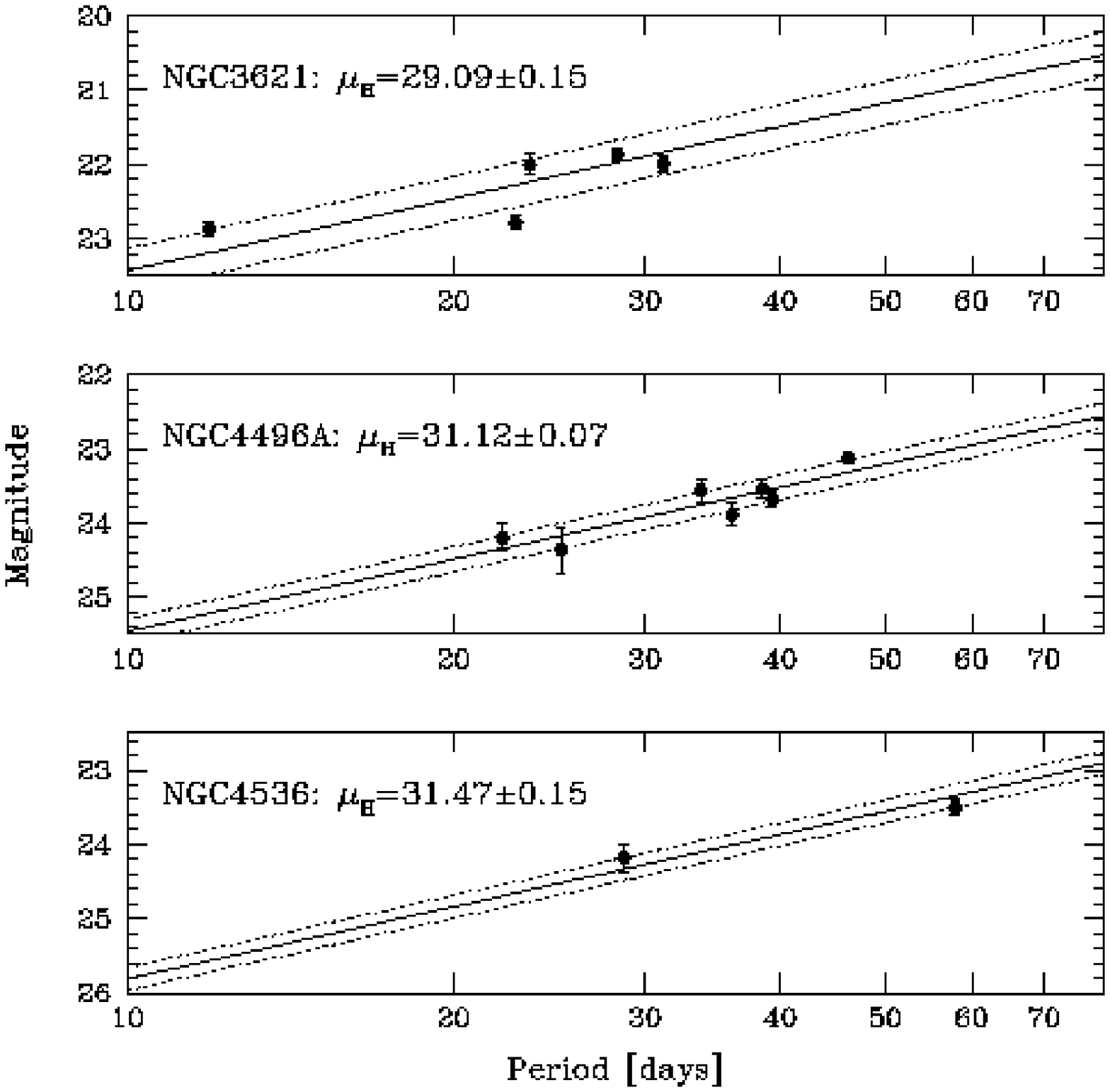}
\caption{(f) (continued)}
\end{figure}

\clearpage

\begin{figure}
\plotone{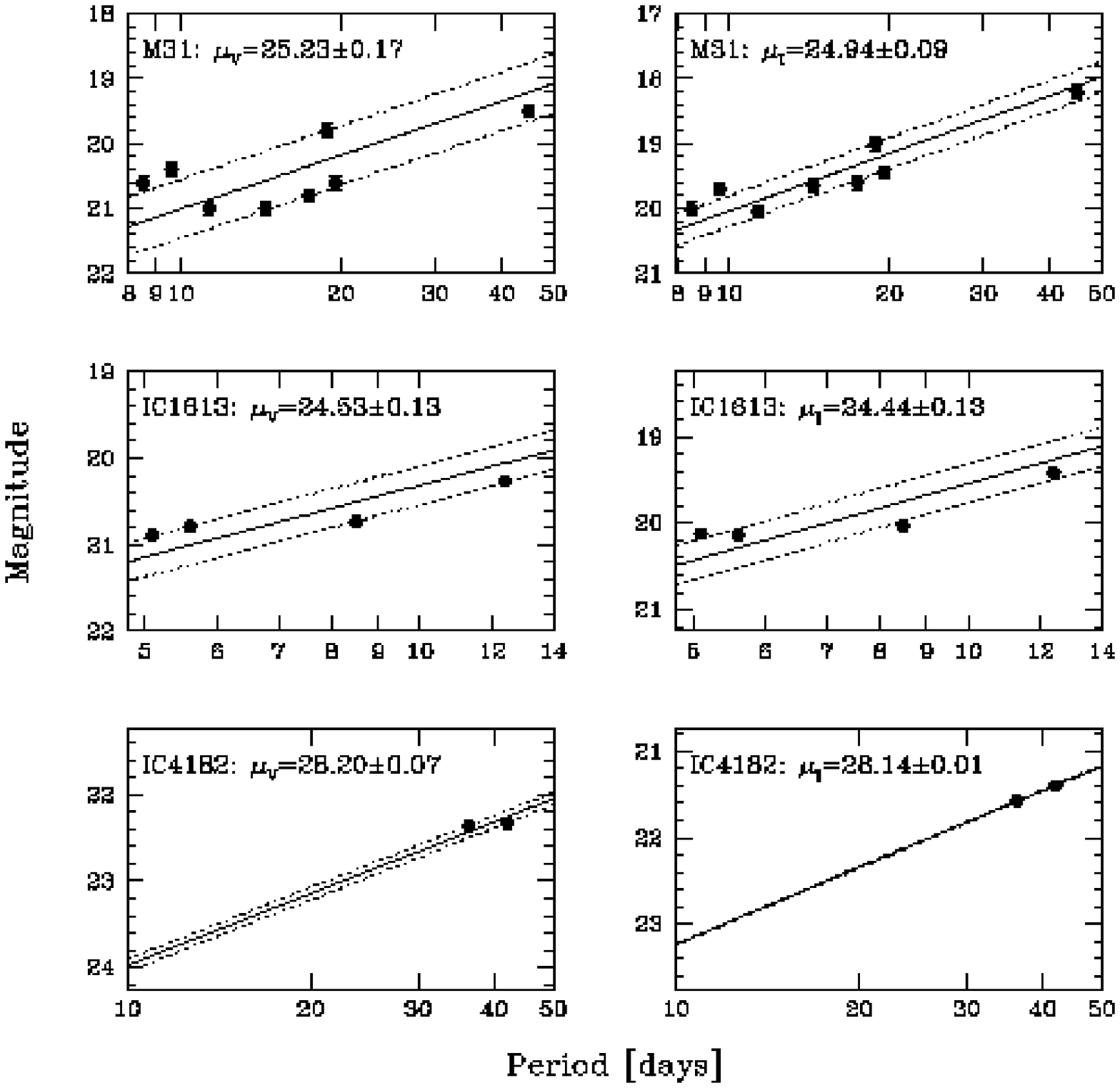}
\caption{(a) $V$ and $I$-band Period-Luminosity relations for the
fields used in this work. Notation is identical to Figures 8a-f.}
\end{figure}

\clearpage

\setcounter{figure}{8}

\begin{figure}
\plotone{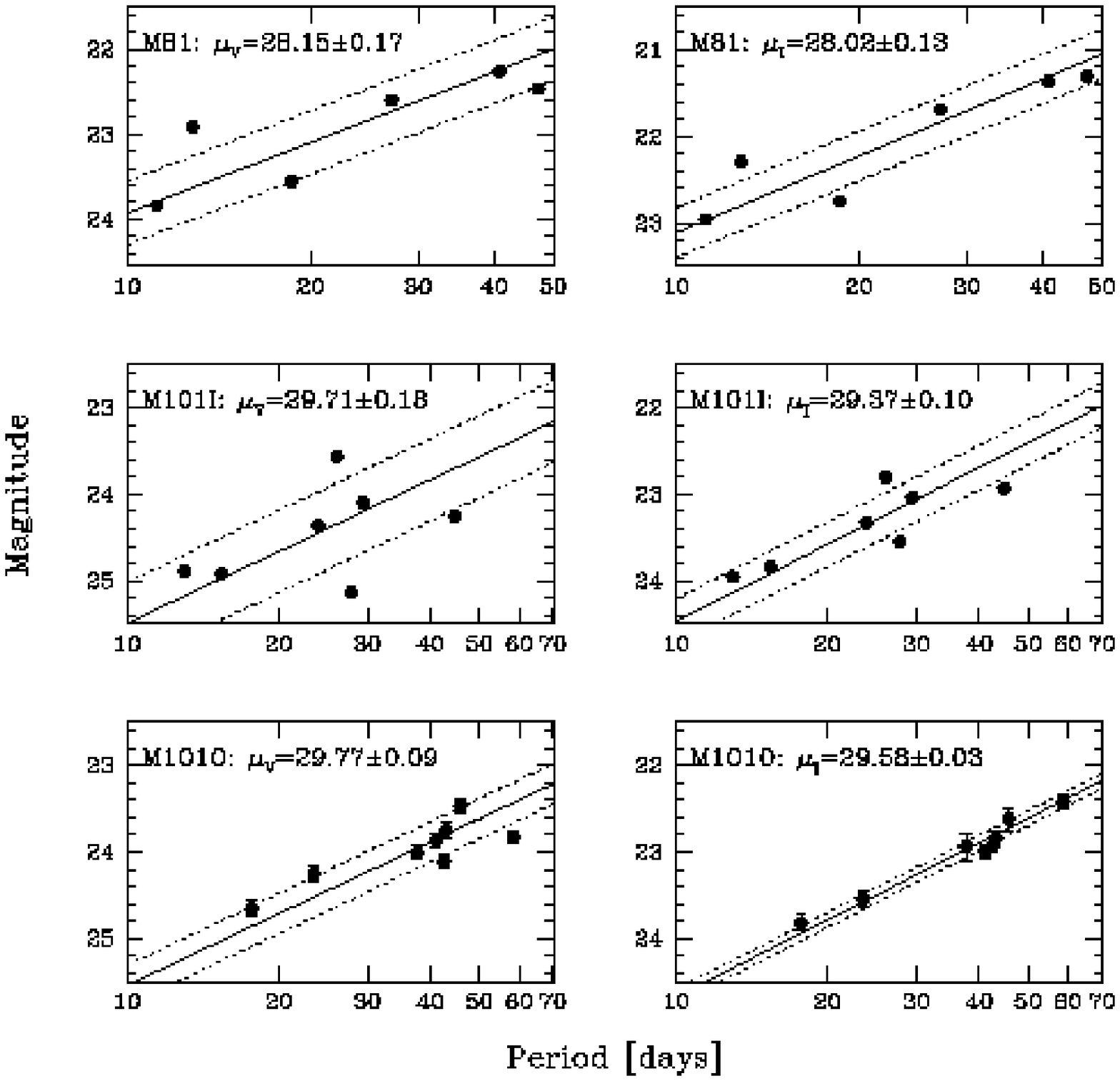}
\caption{(b) (continued)}
\end{figure}

\clearpage

\setcounter{figure}{8}

\begin{figure}
\plotone{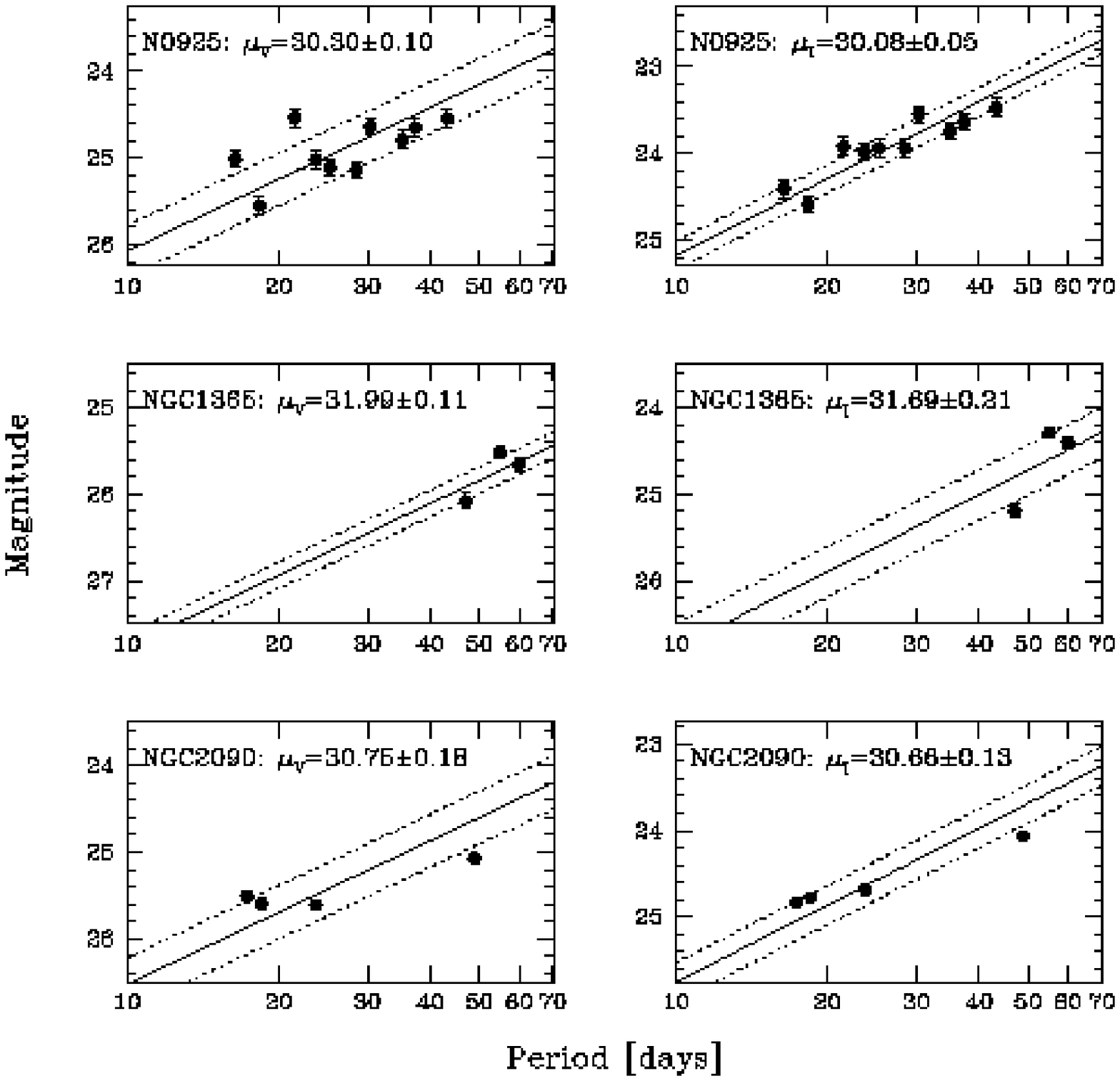}
\caption{(c) (continued)}
\end{figure}

\clearpage

\setcounter{figure}{8}

\begin{figure}
\plotone{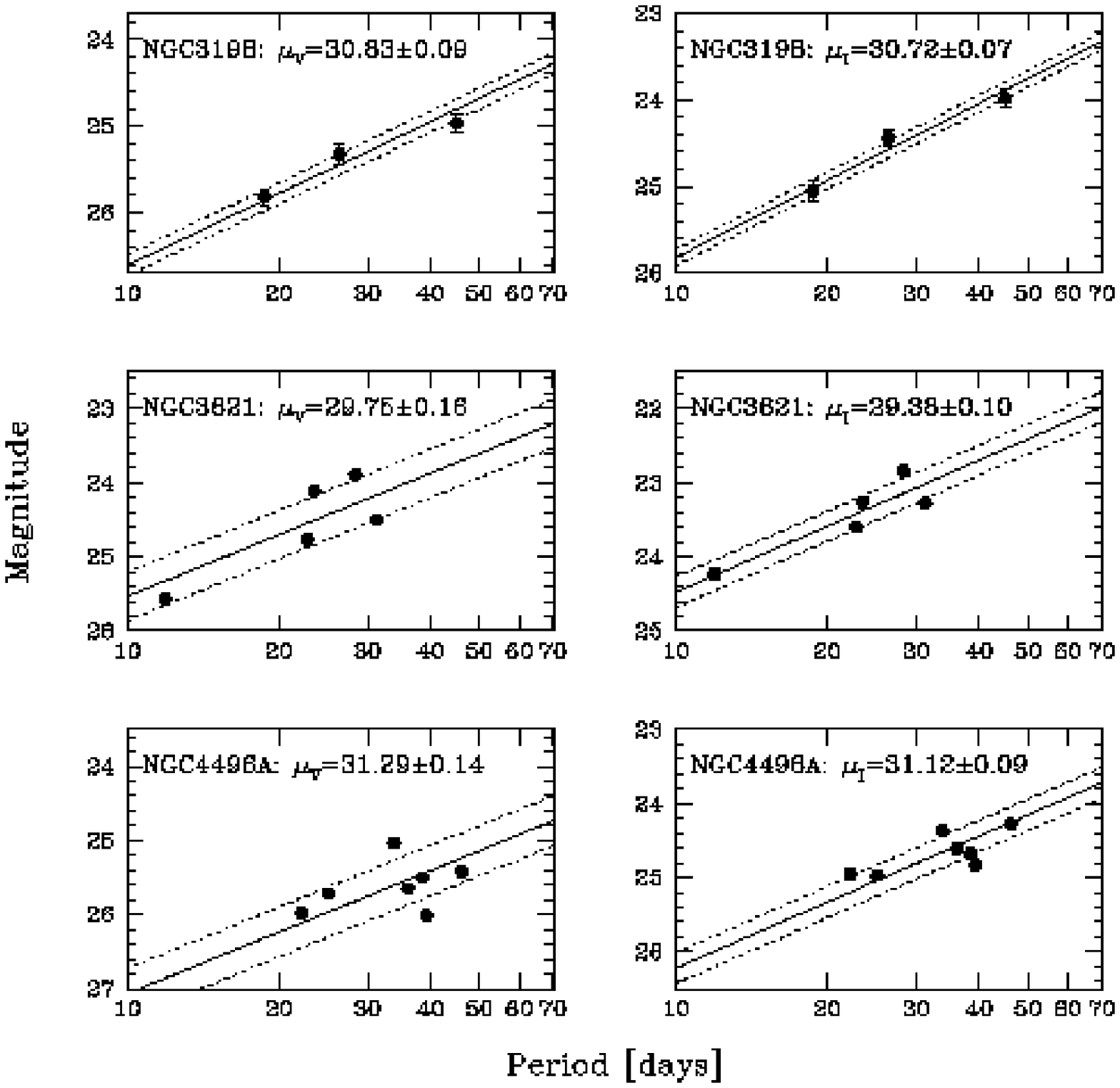}
\caption{(d) (continued)}
\end{figure}

\clearpage

\setcounter{figure}{8}

\begin{figure}
\plotone{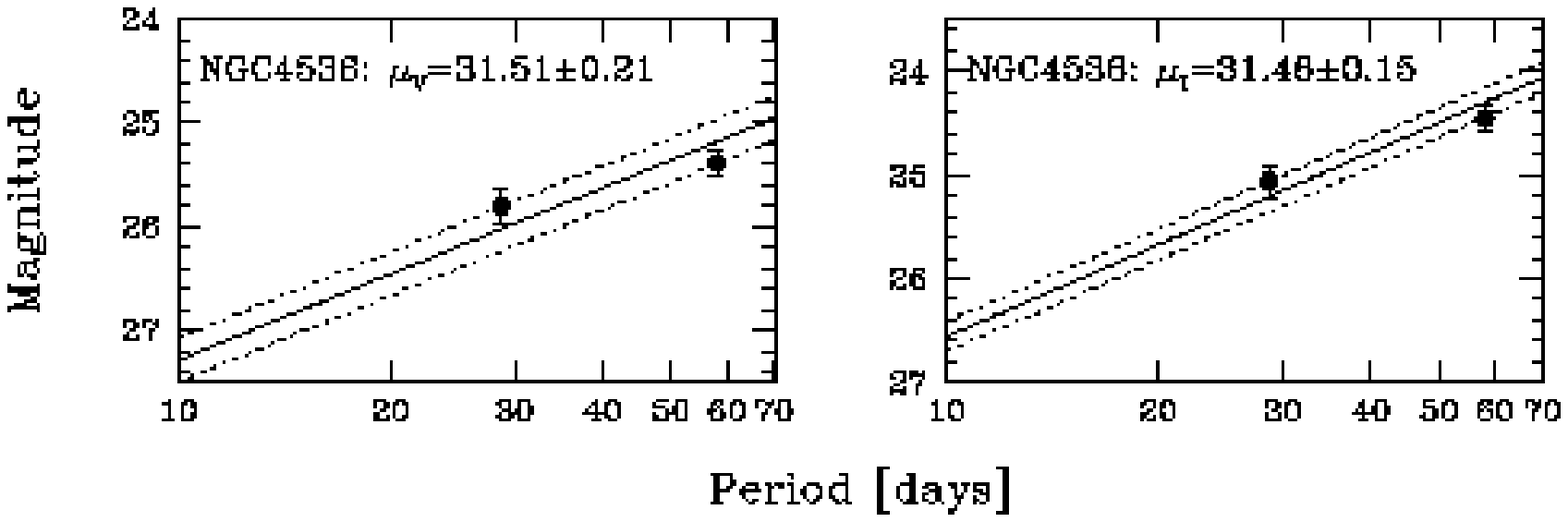}
\caption{(e) (continued)}
\end{figure}

\clearpage

\begin{figure}
\plotone{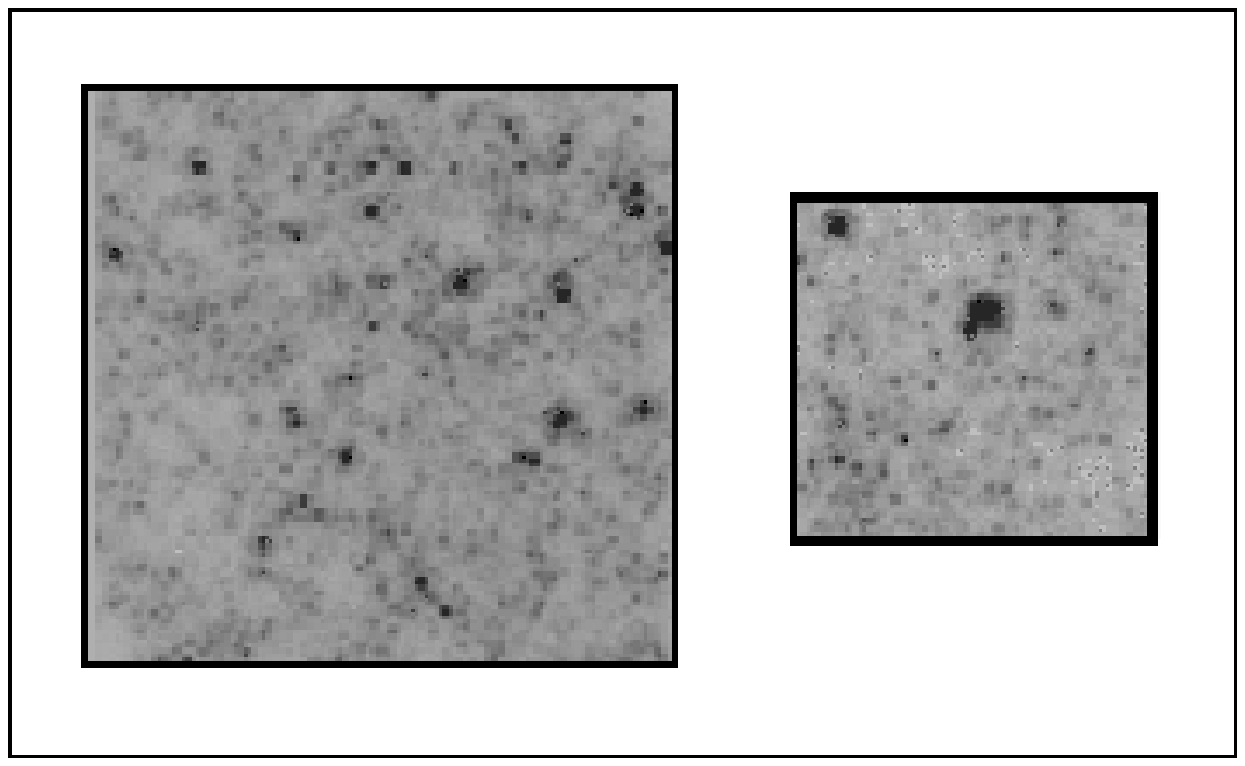}
\caption{Simulated M101 Inner fields, based on our observations of M31 (right)
and M81 (left) fields. These were used to study the effects of blending in our
M101 Inner Cepheids.}
\end{figure}

\clearpage

\begin{figure}
\plotone{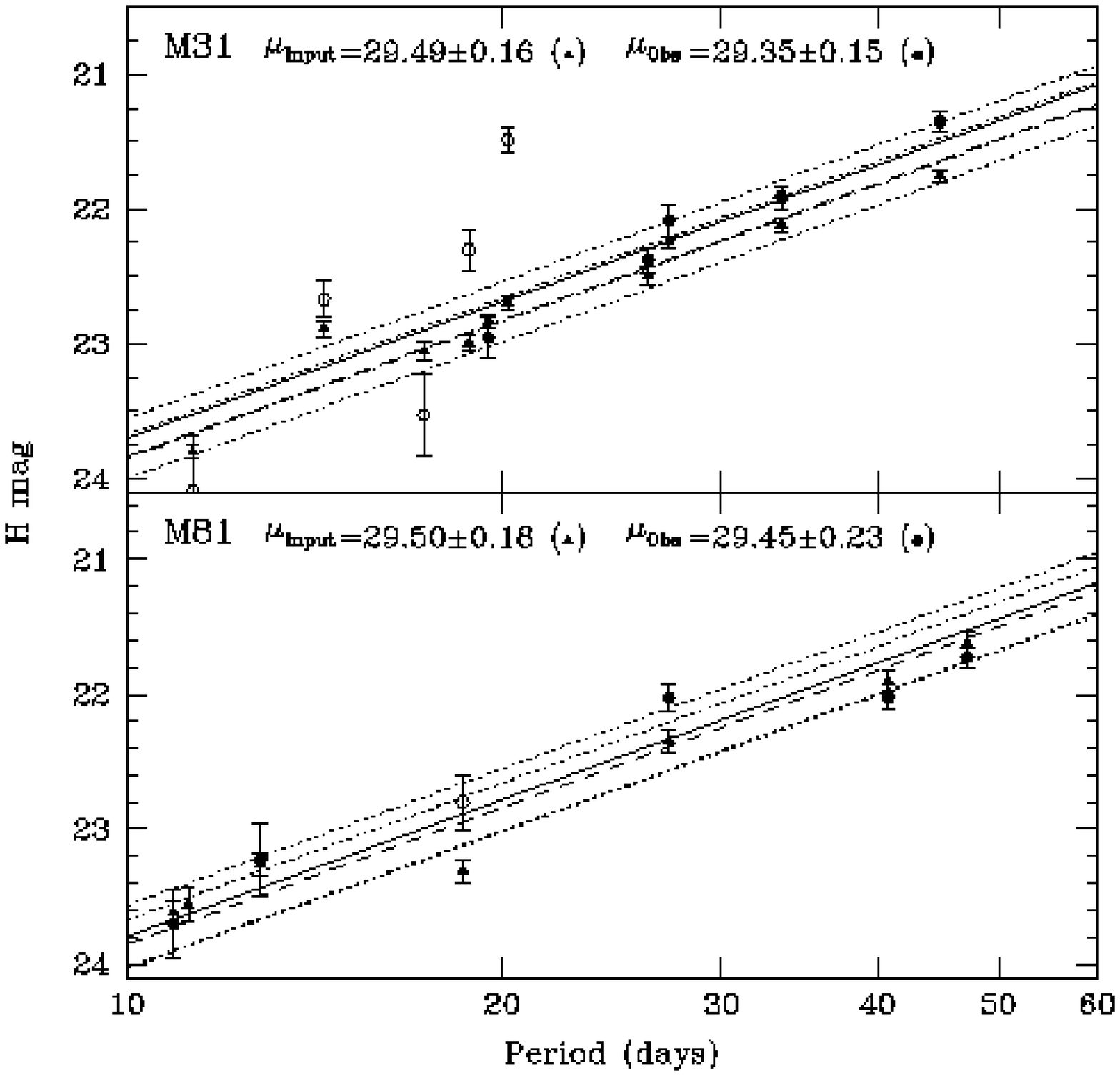}
\caption{Period-Luminosity relations of our simulated M101 Inner fields shown
in Figure 10. Filled triangles indicate the input magnitudes of the Cepheids
(as observed in our original M31 and M81 fields). Circles represent the
recovered magnitudes for the same variables. Filled and open circles are used
to indicate whether the recovered Cepheids would have passed or failed the
color-color test of \S3.1. Dashed and solid lines denote the P-L fits of the
input and recovered magnitudes, respectively.}
\end{figure}

\clearpage

\begin{figure}
\plotone{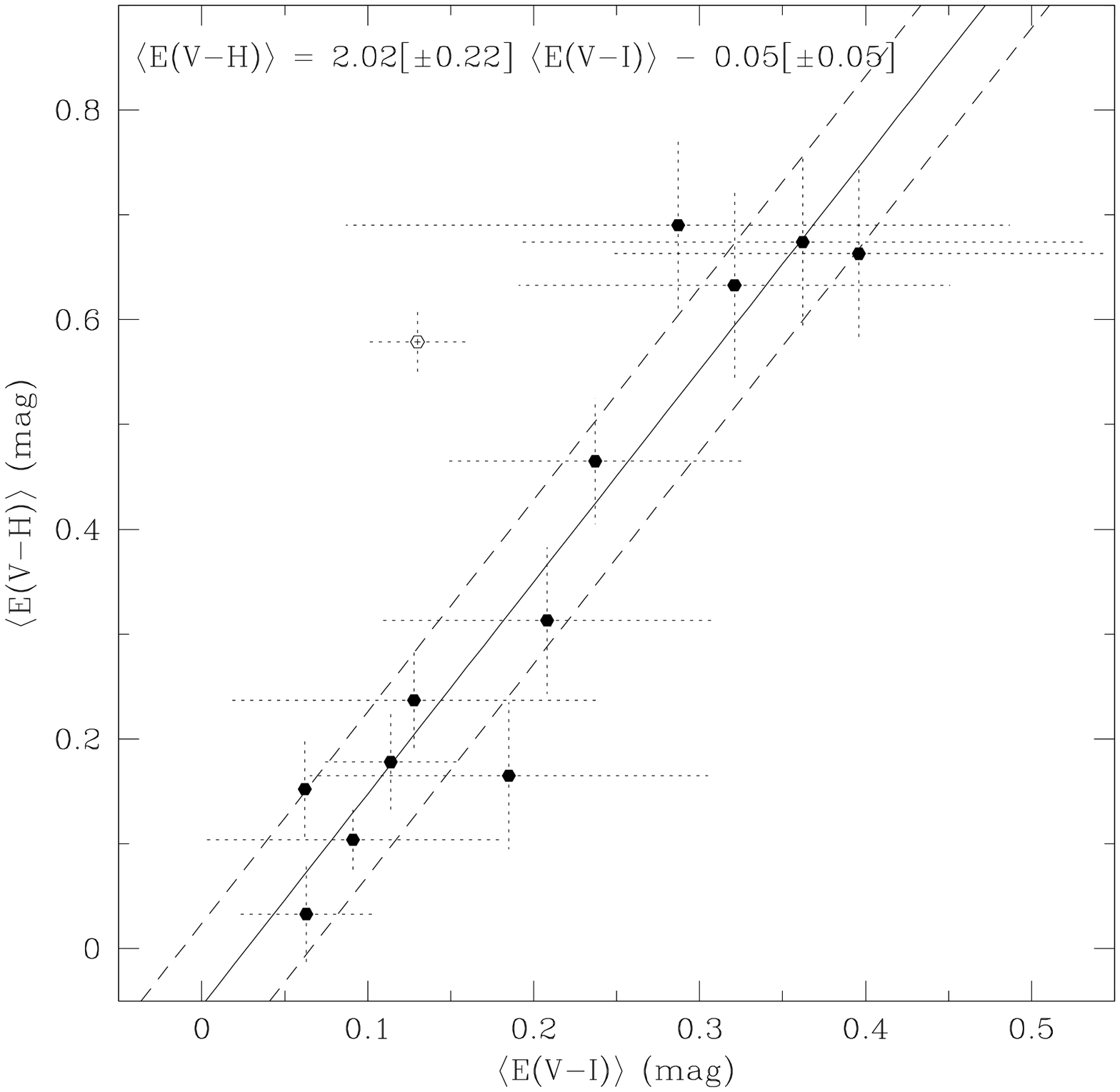}
\caption{A test of the correlation between $\evh$ and $\evi$ for our
fields. Filled circles represent the crowding-corrected color excesses for our
fields. The open circle corresponds to \ng3198, which was rejected from the
fit. The solid line is a least-squares fit to the data, while the dashed lines
indicate the {\it rms} uncertainty of the fit.}
\end{figure}
\end{document}